\documentclass
[aps,prb,superscriptaddress,reprint,showpacs,amssymb,amsmath,citeautoscript,longbibliography]{revtex4-2}

\usepackage[T1]{fontenc}
\usepackage[latin9]{inputenc}
\usepackage{amssymb}

\usepackage{graphicx} %include figure files
\usepackage{dcolumn} %align table columns on decimal point
\usepackage{bm} %bold math

\usepackage{ulem}
\usepackage[]{color}

\begin{document}

\title{Possible multi-orbital ground state in CeCu\textsubscript{2}Si\textsubscript{2}.}

\author{Andrea~Amorese}
 \affiliation{Max Planck Institute for Chemical Physics of Solids, N{\"o}thnitzer Stra{\ss}e 40, 01187 Dresden, Germany}
\author{Andrea~Marino}
 \affiliation{Max Planck Institute for Chemical Physics of Solids, N{\"o}thnitzer Stra{\ss}e 40, 01187 Dresden, Germany}
 \affiliation{Dipartimento di Fisica, Politecnico di Milano, Piazza Leonardo da Vinci 32, I-20133 Milano, Italy}
\author{Martin~Sundermann}
 \affiliation{Institute of Physics II, University of Cologne, Z{\"u}lpicher Stra{\ss}e 77, 50937 Cologne, Germany}
 \affiliation{Max Planck Institute for Chemical Physics of Solids, N{\"o}thnitzer Stra{\ss}e 40, 01187 Dresden, Germany}
\author{Kai~Chen}
\affiliation{Helmholtz-Zentrum Berlin f\"ur Materialien und Energie, Albert-Einstein-Str.15, 12489 Berlin, Germany}
\author{Zhiwei~Hu}
 \affiliation{Max Planck Institute for Chemical Physics of Solids, N{\"o}thnitzer Stra{\ss}e 40, 01187 Dresden, Germany}
 \author{Thomas~Willers}
  \altaffiliation{Kr\"uss GmbH, Borsteler Chaussee 85,  22453 Hamburg, Germany}
 \affiliation{Institute of Physics II, University of Cologne, Z{\"u}lpicher Stra{\ss}e 77, 50937 Cologne, Germany}
\author{Fadi~Choukani}
\affiliation{Synchrotron SOLEIL, L'Orme des Merisiers, Saint-Aubin, 91192 Gif-sur-Yvette, France}
\author{Philippe~Ohresser}
\affiliation{Synchrotron SOLEIL, L'Orme des Merisiers, Saint-Aubin, 91192 Gif-sur-Yvette, France}
\author{Javier Herrero-Martin}
  \affiliation{ALBA Synchrotron Light Source, E-08290 Cerdanyola del Vallès, Barcelona, Spain}
\author{Stefano Agrestini}
  \altaffiliation{present address: Diamond Light Source, Harwell Science and Innovation Campus, Didcot, OX11 0DE, UK}
  \affiliation{Max Planck Institute for Chemical Physics of Solids, N{\"o}thnitzer Stra{\ss}e 40, 01187 Dresden, Germany}
  \affiliation{ALBA Synchrotron Light Source, E-08290 Cerdanyola del Vallès, Barcelona, Spain}
 \author{Chien-Te~Chen}
 \affiliation{National Synchrotron Radiation Research Center (NSRRC), Hsinchu 30077, Taiwan}
\author{Hong-Ji~Lin}
 \affiliation{National Synchrotron Radiation Research Center (NSRRC), Hsinchu 30077, Taiwan}
\author{Maurits~W.~Haverkort}
  \affiliation{Institute for Theoretical Physics, Heidelberg University, Philosophenweg 19, 69120 Heidelberg, Germany}
\author{Silvia~Seiro}
  \altaffiliation{present address: Leibniz IFW Dresden, Helmholtzstr. 20, 01069 Dresden, Germany}	
  \affiliation{Max Planck Institute for Chemical Physics of Solids, N{\"o}thnitzer Stra{\ss}e 40, 01187 Dresden, Germany}	
\author{Christoph~Geibel}
  \affiliation{Max Planck Institute for Chemical Physics of Solids, N{\"o}thnitzer Stra{\ss}e 40, 01187 Dresden, Germany}	
\author{Frank~Steglich}
  \affiliation{Max Planck Institute for Chemical Physics of Solids, N{\"o}thnitzer Stra{\ss}e 40, 01187 Dresden, Germany}
\author{Liu~Hao~Tjeng}
	\affiliation{Max Planck Institute for Chemical Physics of Solids, N{\"o}thnitzer Stra{\ss}e 40, 01187 Dresden, Germany}
\author{Gertrud~Zwicknagl}
  \affiliation{Institute for Mathematical Physics, Technische Universit{\"a}t  Braunschweig, D-38106 Braunschweig, Germany}
\author{Andrea~Severing}
  \affiliation{Institute of Physics II, University of Cologne, Z{\"u}lpicher Stra{\ss}e 77, 50937 Cologne, Germany}
  \affiliation{Max Planck Institute for Chemical Physics of Solids, N{\"o}thnitzer Stra{\ss}e 40, 01187 Dresden, Germany}	
\date{\today}

\begin{abstract}
The crystal-field ground state wave function of CeCu$_2$Si$_2$ has been investigated with linear polarized $M$-edge x-ray absorption spectroscopy from  250\,mK to 250\,K, thus covering the superconducting ($T_{\text{c}}$\,=\,0.6\,K), the Kondo ($T_{\text{K}}$\,$\approx$\,20\,K) as well as the Curie-Weiss regime. The comparison with full-multiplet calculations shows that the temperature dependence of the experimental linear dichroism is well explained with a $\Gamma_7^{(1)}$ crystal-field ground-state and the thermal population of excited states at around 30\,meV. The crystal-field scheme does not change throughout the entire temperature range thus making the scenario of orbital switching unlikely. Spectroscopic evidence for the presence of the Ce 4$f^0$ configuration in the ground state is consistent with the possibility for a multi-orbital character of the ground state. We estimate from the Kondo temperature and crystal-field splitting energies that several percents of the higher lying $\Gamma_6$ state and $\Gamma_7^{(2)}$ crystal-field states are mixed into the primarily $\Gamma_7^{(1)}$ ground state. This estimate is also supported by re-normalized band-structure calculations that uses the experimentally determined crystal-field scheme. 

\end{abstract}

\maketitle

\section{Introduction}
Heavy fermion compounds are $f$ electron systems where, at low temperatures, the hybridization of localized 4$f$ or 5$f$ and conduction electrons ($cf$-hybridization) forms an entangled ground state with quasiparticles that can have effective masses up to three orders of magnitude larger than the free electron mass\,\cite{Coleman2007,Coleman2015}. The $cf$-hybridization goes along with a certain delocalization of the $f$ electrons and depending on the degree of delocalization magnetic order, unconventional superconductivity or intermediate valence occurs. Here superconductivity usually occurs in the vicinity of the quantum critical point where the magnetic order transitions are suppressed to zero Kelvin\,\cite{Hilbert2007,Wirth2016}. In the heavy fermion compound CeCu$_2$Si$_2$, a material with a Kondo temperature of $T_{\text{K}}$\,$\approx$\,20\,K\,\cite{Sun2013,Steglich2016}, unconventional superconductivity was observed for the first time\,\cite{Steglich1979} opening up an entire new field of research. Superconductivity in CeCu$_2$Si$_2$ appears at ambient pressure but also in a wider range of applied pressures where two superconducting domes have been observed with maxima at 0.45\,GPa ($T_{\text{c}}$\,=\,0.6\,K) and 4.5\,GPa ($T_{\text{c}}$\,=\,2\,K) in the pressure($P$)/temperature($T$) phase diagram\,\cite{Thomas1993}. The substitution of Si by the larger Ge separates the two domes\,\cite{Yuan2003} suggesting the two superconducting phases may be of different origin. 

The ambient or low pressure superconducting phase is close to antiferromagnetism; small changes in the Si stoichiometry lead to an antiferromagnetic ground state\,\cite{Stockert2010}. It is therefore likely that spin fluctuations are responsible for the formation of Cooper pairs and there is strong evidence for the $d$-wave character of this superconducting phase\,\cite{Eremin2008,Stockert2011,Pang2018}. For the high pressure superconducting phase, however, valence fluctuations were proposed to provide the pairing mechanism\,\cite{Holmes2007} but so far a valence transition at applied pressure has not been experimentally confirmed\,\cite{Rueff2011}. 

The \textit{d}-wave character of the ambient pressure superconductivity in CeCu$_2$Si$_2$ has been contested recently\,\cite{Kittaka2014,Yamashitae2017,Takenaka2017} and here the determination of the crystal-field wave functions of the crystal-field split  Hund's rule ground state has become important. Is the ground state a \textit{multiorbital} state as suggested by Ref.\,\onlinecite{Emilian2019}?  Or does an orbital switching as function of temperature take place as suggested in another scenario to model the double dome structure of the two superconducting phases\,\cite{Pourovskii2014}?  For further clarification it is therefore indispensable to revisit spectroscopically the crystal-field problem of CeCu$_2$Si$_2$ and its temperature dependence. 

We recall the $J$=\,5/2 Hund's rule ground state of Ce in CeCu$_2$Si$_2$ is split by the tetragonal crystal field into two $\Gamma_7^{1,2}$ and one $\Gamma_6$ Kramers doublet which can be written in $J_z$ representation with 0\,$\leq$\,$\left|\alpha\right|$\,$\leq$\,1:  
\begin{eqnarray}
\Gamma_7^{(1)}&=&\alpha|\pm 5/2\rangle + \sqrt{(1-\alpha^2)}|\mp 3/2\rangle\\
\Gamma_7^{(2)}&=&\sqrt{(1-\alpha^2)}|\pm 5/2\rangle - \alpha|\mp 3/2\rangle\\ 
\Gamma_6  &=&|\pm 1/2 \rangle 
\end{eqnarray}

\begin{figure}[h]
    \includegraphics[width=0.8\columnwidth]{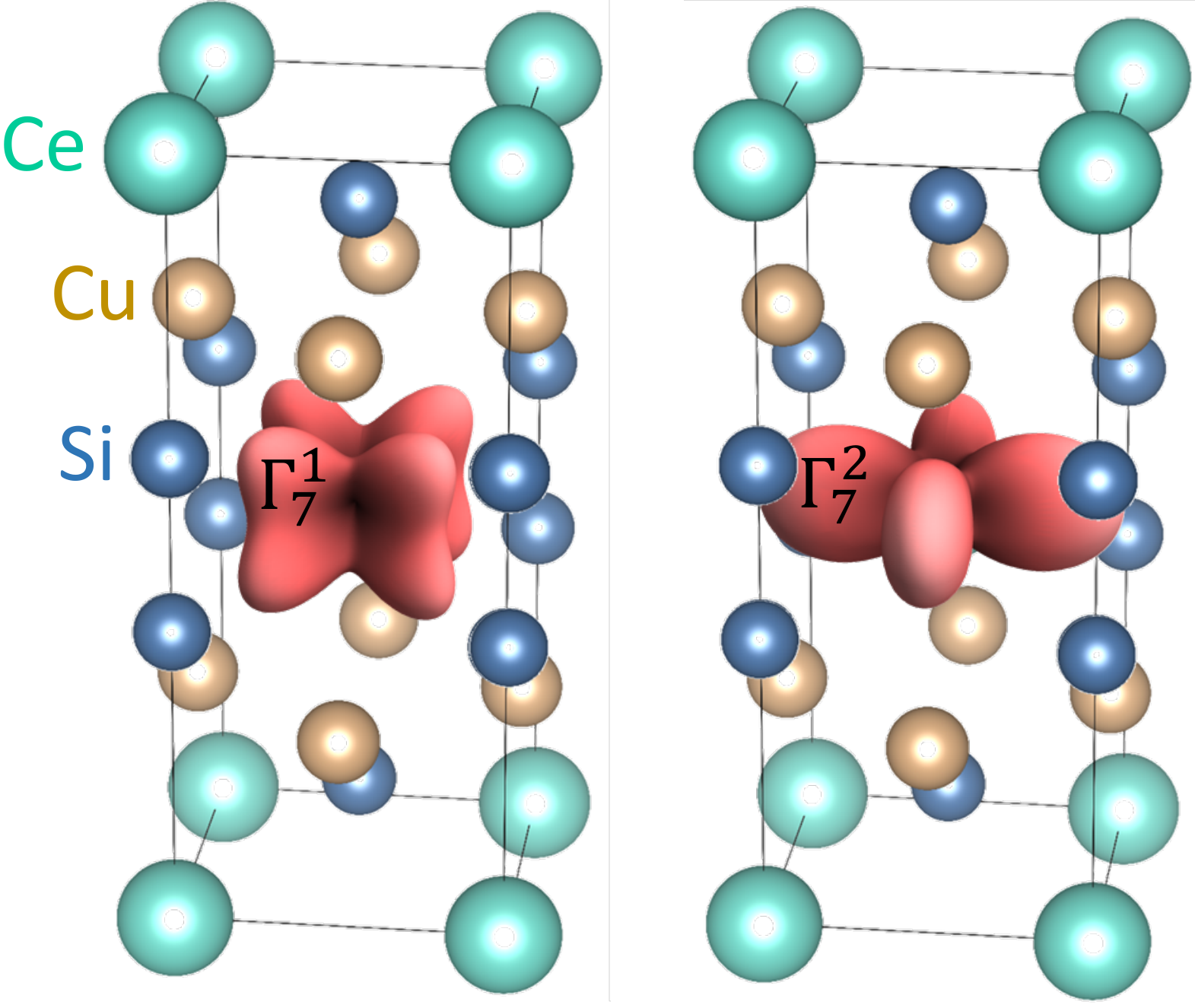}
    \caption{ThCr$_2$Si$_2$ structure of CeCu$_2$Si$_2$ with Ce $\Gamma_7^{(1)}$ (left) and $\Gamma_7^{(2)}$ orbital (right) at the body center of the unit cell. The aspect ratios of the orbitals correspond to eq.(1) and (2) with $\alpha$\,=\,0.59 and the orientation to $\alpha$\,$<$\,0.}
    \label{unit_cell}
\end{figure}

Here the $\Gamma_7^{1,2}$ orbital states distinguish each other in their $J_z$ admixture and orientation within the unit cell; for $\alpha$\,$>$\,0 the lobes of the angular distribution of the $\Gamma_7^{(1)}$ state points along [100] and for $\alpha$\,$<$\,0 along [110] (see Fig.\,\ref{unit_cell}), for $\alpha$\,=\,0 i.e. in case of a pure $J_z$ state the orbital has full rotational symmetry around the [001] axis. Goremychkin \textsl{et al.} found with inelastic neutron scattering two strongly broadened and almost degenerate crystal-field excitations at about 30\,meV\,\cite{Goremychkin1993}. In the Goremychkin neutron experiment the crystal-field wave functions were, however, determined from the anisotropy of the static susceptibility $\chi_{stat}$ which is well described with a $\Gamma_7^{(1)}$ ground state with $\left|\alpha\right|$\,=\,0.88. Note, inelastic neutron scattering cannot determine the sign of $\alpha$ in the wave function since it is dipole limited. Non-resonant inelastic x-ray scattering (NIXS) overcomes this dipole limitation\,\cite{Haverkort2007,Gordon2008,Gordon2009} and Willers \textsl{et al.} found that $\alpha$ in $\Gamma_7^{(1)}$ is negative at 20\,K i.e. the $\Gamma_7^{(1)}$ with its lobes along the (110) direction forms the ground state at 20\,K\,\cite{Willers2012a}. The NIXS experiment was performed well above $T_{\text{K}}$ and $T_{\text{c}}$ without looking explicitly at the $J_z$ admixture ($ac$ anisotropy). Hence, till today, there is no spectroscopic information available about the $J_z$ admixture of the ground-state wave function of CeCu$_2$Si$_2$, nor about the possibility of a $cf$-hybridization induced multiorbital ground state\,\cite{Emilian2019} or orbital reoccupation\,\cite{Pourovskii2014}. 

The present work addresses this lack of information. We set up an experiment with the aim to investigate the crystal-field wave functions of the ground state, below $T_{\text{K}}$ and $T_{\text{c}}$, and well above, looking also for any changes in the orbital occupation that cannot be explained with the Boltzman-type thermal occupation of excited crystal-field states.   

\section{Method}
X-ray absorption spectroscopy (XAS) is an element specific probe for valence, spin and orbital degrees of freedom\,\cite{Tanaka1994,Thole1997,deGroot2008}. In particular, the linear dichroism (LD) of linear polarized XAS (XLD) at the rare earth $M_{4,5}$ edges (3$d^{10}$4$f^1$\,$\rightarrow$\,3$d^{9}$4$f^2$) measures the anisotropy of the 4$f^1$ wave function of the crystal-field field split Hund's rule ground state of Ce$^{3+}$ ($J$\,=\,5/2) with unprecedented accuracy\,\cite{Hansmann2008,Willers2010,Willers2012,Strigari2012,Willers2015}. XAS is element specific and the signal to background ratio is very good. XLD probes specifically the ground-state symmetry when working at low $T$. Excited crystal-field states contribute via thermal occupation.  

XLD is based on dipole selection rules and each $J_z$ state exhibits its own specific directional dependence\,\cite{Hansmann2008} resulting in a specific dichroism LD$_{J_z}$ that relate to each other as LD$_{5/2}$\,=\,$-5$\,LD$_{3/2}$\,=\,$-1.25$\,LD$_{1/2}$. Although the data analysis was performed with a full multiplet calculation (see below), the analysis becomes more intuitive when expressing the LD of each crystal-field state in terms of incoherent sums of the individual LD$_{J_z}$. This is only possible when the rotational symmetry is higher than twofold and as long as the crystal-field splitting is small with respect to the spin orbit splitting\,\cite{Hansmann2008}. Both are fulfilled for CeCu$_2$Si$_2$ so that we can write for the LD of the $\Gamma_7^{(1)}$ state: 
\begin{eqnarray}
LD_{\Gamma_7^{(1)}}&=&\alpha^2LD_{5/2} + (1-\alpha^2)LD_{3/2} 
\end{eqnarray}
The LD$_{\Gamma_7^2}$ of the $\Gamma_7^{(2)}$ state can be written accordingly. This little exercise demonstrates that XLD is sensitive to the square of $\alpha$ and therefore not to its sign. When excited states get populated with rising temperature the total LD($T$) signal is the superposition of the individual LDs of each crystal-field state, weighted by thermal occupation.

A prove for the existence of $cf$-hybridization is the presence of a satellite in the $M_{4,5}$-edge spectra. It arises from the $cf$-hybridization induced mixture of the 4$f^1$\,(Ce$^{3+}$\,$J$=5/2) and 4$f^0$\,Ce$^{4+}$\,$J$=0) configurations. The core hole splits up the final states that can be reached from the quantum mechanical entangled ground state given by $\sqrt{n_1}$\,$f^1$\,+\,$\sqrt{n_0}$\,$f^0$\,\cite{Gunnarsson1983} since it acts differently on different configurations. Here $n_1$ and $n_0$ label the weights of the two configurations. The satellite indicating the presence of 4$f^0$ in the ground state shows up at the high energy tail of the 4$f^1$ absorption lines; here we are always referring to the configurations of the initial state\,\cite{Hansmann2008}. These satellites can be used for measuring relative changes of the 4$f$ shell occupation with temperature\,\cite{Willers2012}. 

\section{Experiment and simulations}
The XLD experiment was performed on well characterized superconducting CeCu$_2$Si$_2$ single crystals\,\cite{Seiro2010} at the DEIMOS beamline at synchrotron SOLEIL in France\,\cite{Joly2014} between 0.25 and 5\,K, at the BOREAS beamline at synchrotron ALBA in Spain\,\cite{Barla2016} between 3.2 and 150\,K and the DRAGON beamline of the NSRRC in Taiwan between 100 and 250\,K. The energy resolution at the Ce $M_{4,5}$ edge at $h\nu$\,$\approx$\,870-910\,eV was about 0.4\,eV. The DEIMOS beamline is quite unique in the world since its cryomagnet is equipped with an insert for cooling to 250\,mK in ultra high vacuum (UHV)\,\cite{Kappler2018}. The temperature stability is within a few percent and the temperature difference between thermocouple and sample surface amounts to 50 to 100\,mK depending on thermal contact. The BOREAS beamline has the advantage of a reference sample in the beam so that small drifts in energy can be easily corrected.  At all beamlines the samples were cleaved \textit{in situ} in ultra high vacuum and then transferred to a main chamber (10$^{-10}$\,mbar) where the signal was recorded in the total-electron-yield (TEY) mode by measuring the drain current. The cleaved $ac$ surface was perpendicular to the Poynting vector, with $c$ being the fourfold tetragonal axis, so that data could be taken with the electric field vector $\vec{E}\|c$ and $\vec{E} \bot c$. This was achieved at the DEIMOS and BOREAS beamline by changing the polarization of the light impinging on the sample. At the DRAGON beamline the polarization cannot be changed and, instead, the sample is turned to achieve the polarization parallel and perpendicular to the $c$-axis. All samples were aligned with the Laue method prior to the experiment. 

The data were normalized to the integrated intensity of the experimental isotropic spectra, constructed  as $I_{iso}$\,=\,($I_{\vec{E} \| c}$\,$+$\,2\,$I_{\vec{E} \bot c}$)/3, and compared with simulations obtained with the full multiplet code $Quanty$\,\cite{Haverkort2016}. The atomic parameters for the 4$f$-4$f$ and 3$d$-4$f$ Coulomb interactions were calculated with the Cowan code\,\cite{Cowan1981} and reduced by about 21\% and 39\%, respectively, to account for configuration interaction effects that are not included in the Hartee-Fock scheme. The reduction factors are determined by optimizing the XAS simulation to $I_{iso}$ without taking into account the crystal-field splitting of the Hund's rule ground state. Configuration interaction effects are not considered. For obtaining the best crystal-field description of the linear polarized data the linear dicroism LD\,=\,$I_{\vec{E} \| c}$\,$-$\,$I_{\vec{E} \bot c}$ was  fitted because the background due to the edge jump as well as the 4$f^0$ satellite do not show any dichroism and so they cancel out in the LD. 

\section{Results}

\begin{figure}[t]
    \includegraphics[width=0.94\columnwidth]{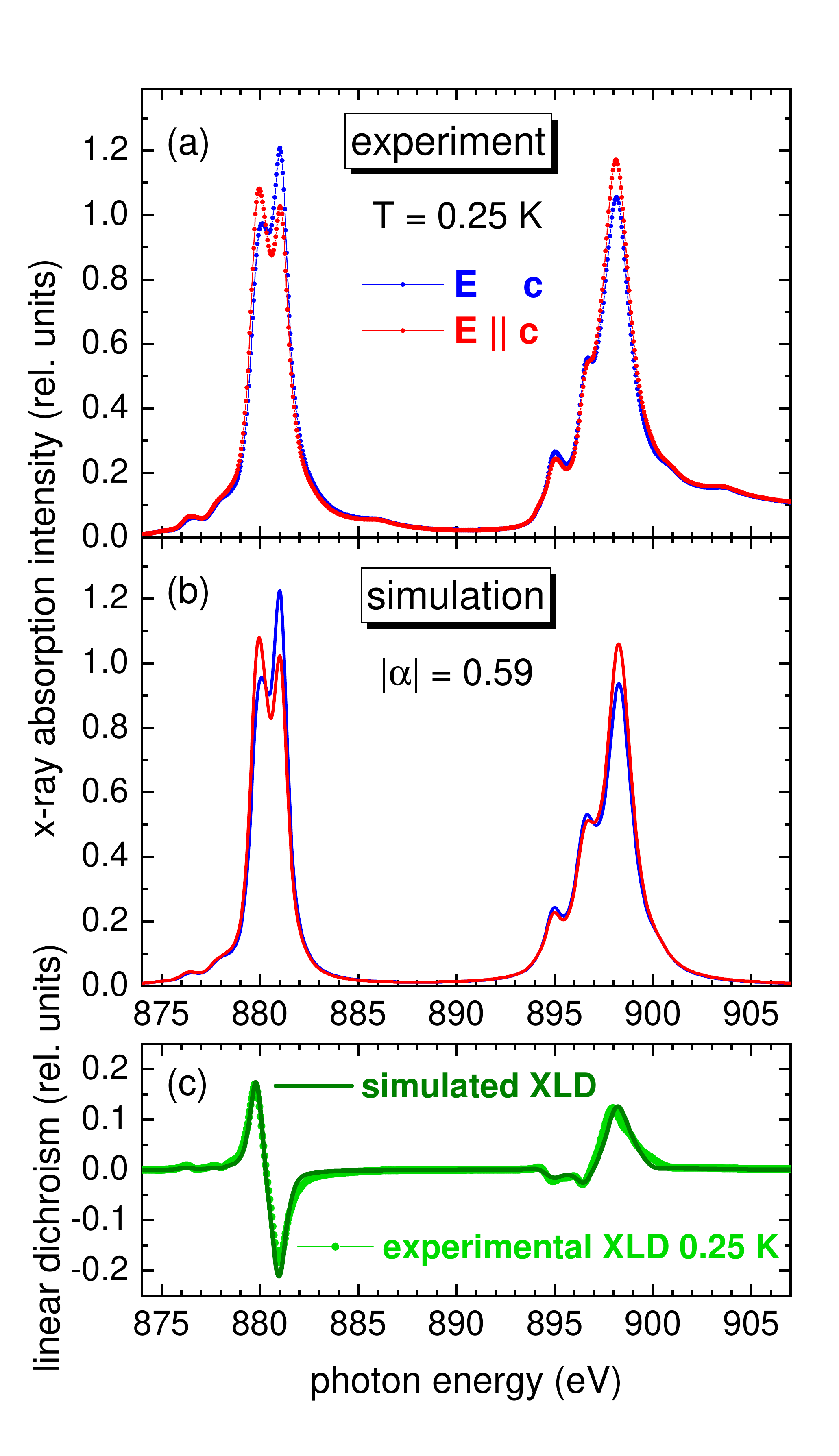}
    \caption{(color online) (a) Experimental linear polarized x-ray absorption data $I_{\vec{E} \| c}$ and $I_{\vec{E} \bot c}$ of CeCu$_2$SI$_2$ at 250\,mK, (b) simulation with a $\Gamma_7^{(1)}$ ground-state wave function (see eq.(1)) and $\left|\alpha\right|$\,=\,0.59, and (c) the linear dichroism LD\,=\,$I_{\vec{E} \| c}$\,$-$\,$I_{\vec{E} \bot c}$ demonstrating the excellent agreement between data (light green dots) and simulation (dark green line).}
    \label{XAS_250mK}
\end{figure}

Figure\,\ref{XAS_250mK}\,(a) shows the XLD data of CeCu$_2$Si$_2$ at 0.25\,K and panel (b) the simulation based on a single ion crystal-field model with a ground state $\Gamma_7^{(1)}$ (eq.\,(1)) with $\alpha$\,=\, $\left|0.59\right|$\,$\pm$\,0.1. The comparison of the experimental and simulated dichroism LD\,=\,$I_{\vec{E} \| c}$\,$-$\,$I_{\vec{E} \bot c}$ in Fig.\,\ref{XAS_250mK}\,(c) establishes how well the data are reproduced. 

\begin{figure}[h]
    \includegraphics[width=0.94\columnwidth]{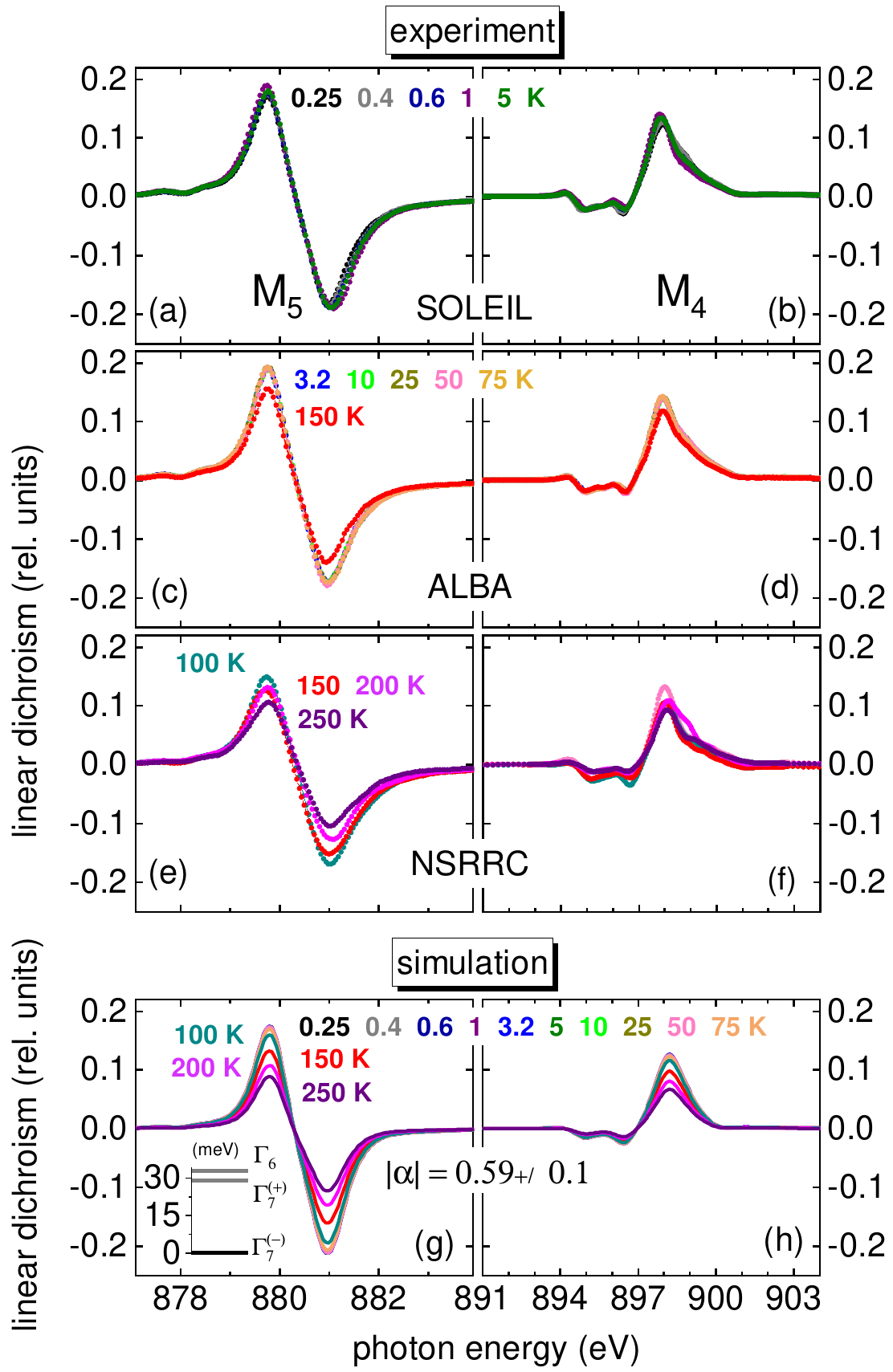}
    \caption{(color online) Experimental linear dichroism LD\,=\,$I_{\vec{E} \| c}$\,$-$\,$I_{\vec{E} \bot c}$ at the $M_5$ and $M_4$ edges for $T$\,=\,0.25 to 5\,K (SOLEIL data) in panel (a) and (b) and for $T$\,=\,3.2 to 150\,K (ALBA data) in panel (c) and (d), and $T$\,=\,100 to 250\,K (NSRRC data) in panel (e) and (f). Panel (g) and (h) shows the full multiplet simulations taking into account the thermal population of crystal-field states at around 30\,meV.}
    \label{T_dep}
\end{figure}

The temperature dependence of the LD is displayed in Fig.\,\ref{T_dep}\,(a)-(d); for $T$\,=\,0.25 to 5\,K in panel (a) and (b), for $T$\,=\,3.2 to 150\,K in panel (c) and (d), and for $T$\,=\,100 to 250\,K in panel (e) and (f). Within the accuracy of the experiment, there is no change in the LD up to 75\,K. At 150\,K, however, the LD decreases a fair amount. Figure\,\ref{T_dep}\,(g)\,and\,(h) show the simulated LD on the basis of a $\Gamma_7^{(1)}$ ground-state wave function with $\alpha$\,=\, $\left|0.59\right|$ and excited states $\Gamma_6$ and $\Gamma_7^{(2)}$ at around 30\,meV: The latter crystal-field splitting energies are from the neutron results in Ref.\,\onlinecite{Goremychkin1993}. The simulations reproduce the experimental data very well. A crystal-field state at e.g. 12\,meV as suggested by Horn \textit{et. al}\,\cite{Horn1981} on the other hand would lead to a much faster reduction of LD and can therefore be excluded. 

\section{Discussion}

The XAS spectra and the LD therein of In CeCu$_2$Si$_2$ have been measured in the wide temperature range from 250\,mK to 250\,K. The data are all well described with the \textit{single-orbital} crystal-field ground state wave function  
\begin{eqnarray}
\Gamma_7^{(1)}&=&-0.59|\pm5/2\rangle + 0.81|\mp3/2\rangle
\end{eqnarray}
and the Boltzman occupation of the excited states at $\approx$30\,meV.  We especially note that there are no noticeable spectral changes while going from the superconducting, via the Kondo, to the Curie-Weiss regime up to 75 K. This strongly suggest that the orbital switching as function of temperature as suggested in Ref.\,\cite{Pourovskii2014} does not take place. Instead, the XAS/LD results point towards a robust static crystal-field scheme. The orbital switching scenario was also discarded for UCu$_2$Ge$_2$ where the in-plane anisotropy of the occupied orbital was measured as function of temperature. In case of an orbital reoccupation of the $\Gamma_7^{(1)}$ and $\Gamma_7^{(2)}$ states, the directional dependence of the NXIS signal should have flipped. This, however, was not observed\,\cite{Rueff2015}.

The question is now whether the ground state is given by just one crystal-field state or whether higher lying crystal-field states contribute. The answer to this question is of relevance for the development of multi-orbital based models\,\cite{Emilian2019} that perhaps could explain why the \textit{d}-wave superconductor CeCu$_2$Si$_2$ may be fully gaped at sufficiently low temperatures \cite{Kittaka2014,Yamashitae2017,Takenaka2017}. 

\begin{figure}[h]
    \includegraphics[width=0.94\columnwidth]{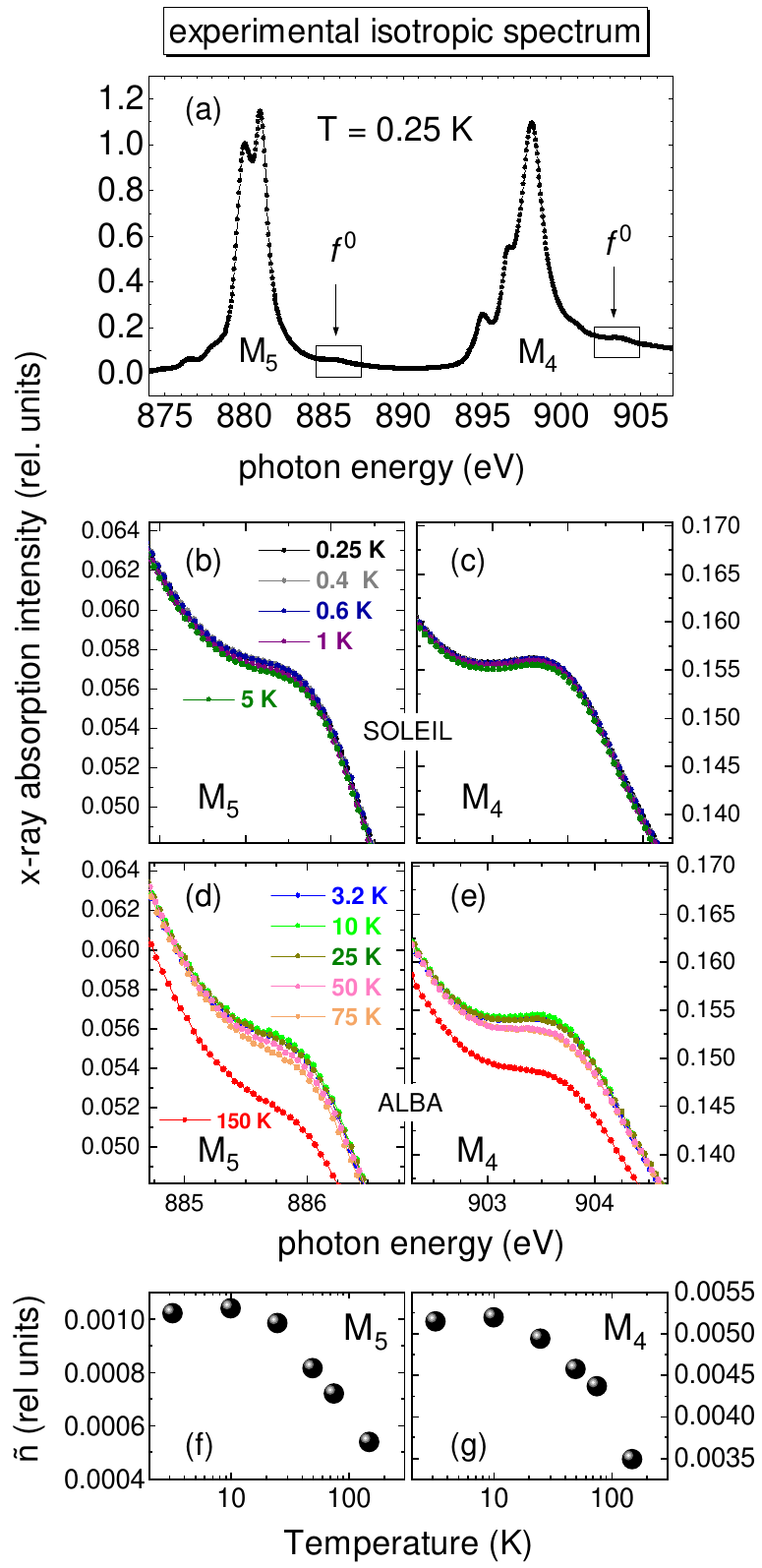}
    \caption{(color online) (a) Isotropic XAS data  $I_{iso}$\,=\,($I_{\vec{E} \| c}$\,$+$\,2\,$I_{\vec{E} \bot c}$)/3 of CeCu$_2$Si$_2$ at 0.25\,K. (b)-(e) $f^0$ satellite on an enlarged scale (see rectangles in panel (a) and (b)) for 0.25 to 5\,K (SOLEIL data) in (a)\,(b) for and 3.2 to 150\,K (ALBA data) in (d)\,(e). (f) and (g) integrated intensity $\tilde{n}$ of $f^0$ satellite in (d) and (e) after subtracting a linear background.}
    \label{isotropic}
\end{figure}

The experimental XAS spectra, see Fig.\,\ref{isotropic}\,(a), exhibit the small but distinct feature at about 886 eV photon energy which can be attributed to the presence of the Ce 4$f^0$ configuration in the ground state\,\cite{Gunnarsson1983}. This in turn signals that the $cf$-hybridization is active and thus also provides a channel for the higher lying crystal-field states to mix into the ground state. The amount of mixing-in is not negligible for systems with high Kondo temperatures such as CeRu$_4$Sn$_6$\,\cite{Sundermann2015} and also CeCoIn$_5$\,\cite{Sundermann2019} where the crystal-field splittings are not much larger than the Kondo energy scale. For CeCu$_2$Si$_2$, however, we expect that the involvement of the higher crystal-field states will be quite small because transport and thermodynamic measurements indicate a Kondo temperature of about 20\,K \cite{Sun2013,Steglich2016} which is small with respect to the crystal-field splitting $\Delta_{\Gamma}$\,$\approx$\,30\,meV\,\cite{Goremychkin1993}. 

In the following we will utilize our spectroscopy to obtain an independent check on the Kondo energy scale by looking at the evolution of the relative 4$f^0$ spectral weight as function of temperature. Panel\,(b)-(e) of Fig.\,\ref{isotropic} display the $f^0$ satellites (see rectangles in panel (a) and (b)) on an enlarged scale. We find that the 4$f^0$ spectral weight remains unchanged up to 5\,K (panel (b) and (c)) within the accuracy of the data, decreases slightly at 50 and 75\,K (panel(d) and (e)), and is strongly reduced at 150\,K. Panel (f) and (g) of Fig.\,\ref{isotropic} show $\tilde{n}$($T$), the integrated intensity of the $f^0$ satellite between $T$\,=\,3.2 and 150\,K after subtracting a linear background. According to a self consistent large-orbital-degeneracy theory by Bickers, Cox and Wilkins\,\cite{Cox1987} the presence of the Kondo effect is reflected in the temperature dependence of $\tilde{n}$($T$). It should show a gradual decrease with temperature followed by a flattening out for $T$\,=\,$\infty$\,$\gg$\,$T_{\text{K}}$. In particular, at $T_{\text{K}}$ it exhibits an inflection point at 1/2 of the difference of its high and low temperature value. This scenario, however, does not take crystal-field effects into account. Therefore, although here we do not observe the inflection point nor flattening out of $\tilde{n}$($T$), possibly because of the population of excited crystal-field states (see Ref.\,\onlinecite{Kummer2018}), we do observe \sout{a strong drop of $\tilde{n}$($T$) not far from } that $\tilde{n}$($T$) starts to drop for $T$\,$>$\,20\,K. We take this as a sign for coming out of the low temperature regime where the Kondo interaction is most effective, and, therefore, interpret the $\tilde{n}$($T$) data in terms of a Kondo temperature of about 20\,K. This value is in good agreement with the the literature value of $T_{\text{K}}$\,=\,20\,K\,\cite{Sun2013,Steglich2016}. We thus find CeCu$_2$Si$_2$ is indeed a material with a relatively low $T_{\text{K}}$.

We now carry out an analysis using a simple approximation scheme for the Anderson impurity Hamiltonian\,\cite{Zwicknagl90a} in order to estimate how much of the higher lying $\Gamma_6$ and $\Gamma_7^{(2)}$ crystal-field states contribute to the primarily $\Gamma_7^{(1)}$ ground state.  The approximate description is valid at temperatures $T\lesssim T_{0}$ where $T_{0}$ is the characteristic temperature of heavy fermion or valence fluctuating systems and the results smoothly reduce to the predictions of the variational treatment\,\cite{Varma1976,Gunnarsson1983}. Assuming a constant density of conduction states $N\left(0\right)$, the occupancies of
the crystal-field split $4f$-states are given by 
\[
n_{f\Gamma\tau}\sim\left(1-n_{f}\right)\left|V_{\Gamma}\right|^{2}N\left(0\right)\frac{1}{k_{B}T_{0}+\Delta_{\Gamma}}
\]
where $\Gamma$ refers to the representation, i. e., $\Gamma$ = $\Gamma_7^{(1)}$, $\Gamma_6$, or $\Gamma_7^{(2)}$, $\tau=\pm$ accounts for the Kramers degeneracy, and $\Delta_{\Gamma}$ is the crystal-field excitation energy relative to the crystal-field ground state. This expression is a straightforward generalization of the formula given in Ref.\,\cite{Gunnarsson1983}. In the systems under consideration, the crystal-field splitting largely exceeds the characteristic energy $\Delta_{\Gamma}\gg k_{B}T_{0}$. The cf hybridization $\left|V_{\Gamma}\right|$ may depend on the symmetry of the crystal-field state $\Gamma$ and the weight of the $f^{0}$-configuration in the ground state is given by 
\[
\left(1-n_{f}\right)=\frac{1}{1+\sum_{\Gamma'}\frac{2\left|V_{\Gamma'}\right|^{2}N\left(0\right)}{k_{B}T_{0}+\Delta_{\Gamma'}}}\]

In CeCu$_{2}$Si$_{2}$, $\left(1-n_{f}\right) \sim0.01-0.03$ and  the dominant contribution to the low-energy states comes from the $\Gamma_7^{1}$ crystal-field ground state. We anticipate, however, a $\Gamma_6$-contribution 
\[
\sum_{\tau=\pm}n_{f\Gamma_{6}\tau}=\frac{\frac{2\left|V_{\Gamma_{6}}\right|^{2}N\left(0\right)}{k_{B}T_{0}+\Delta_{\Gamma_{6}}}}{1+\sum_{\Gamma'}\frac{2\left|V_{\Gamma'}\right|^{2}N\left(0\right)}{k_{B}T_{0}+\Delta_{\Gamma'}}}\sim\frac{\left|V_{\Gamma_{6}}\right|^{2}}{\left|V_{\Gamma_{7}^{(1)}}\right|^{2}}\,\frac{k_{B}T_{0}}{\Delta_{\Gamma_{6}}}\sim\frac{k_{B}T_{0}}{\Delta_{\Gamma_{6}}}
\]
since the hybridization strengths are comparable (see Figure 5 below, renormalized quasiparticle bands). Taking $T_{0}$\,=\,$T_K$\,$\sim$\,20\,K and  $\Delta_{\Gamma_6}$\,$\sim$\,30\,meV, we find about 6\% $\Gamma_6$-contribution. Similarly, with $\Delta_{\Gamma_7^{(2)}}$\,$\sim$\,30\,meV, we may also expect to find about 6\% of $\Gamma_7^{(2)}$-contribution in the ground state. These are small amounts but not negligible, and perhaps sufficient to justify the multi-orbital model\,\cite{Emilian2019} that explains the symmetry of the superconducting gap at very low temperatures.

\begin{figure}[h]
    \includegraphics[width=0.9\columnwidth]{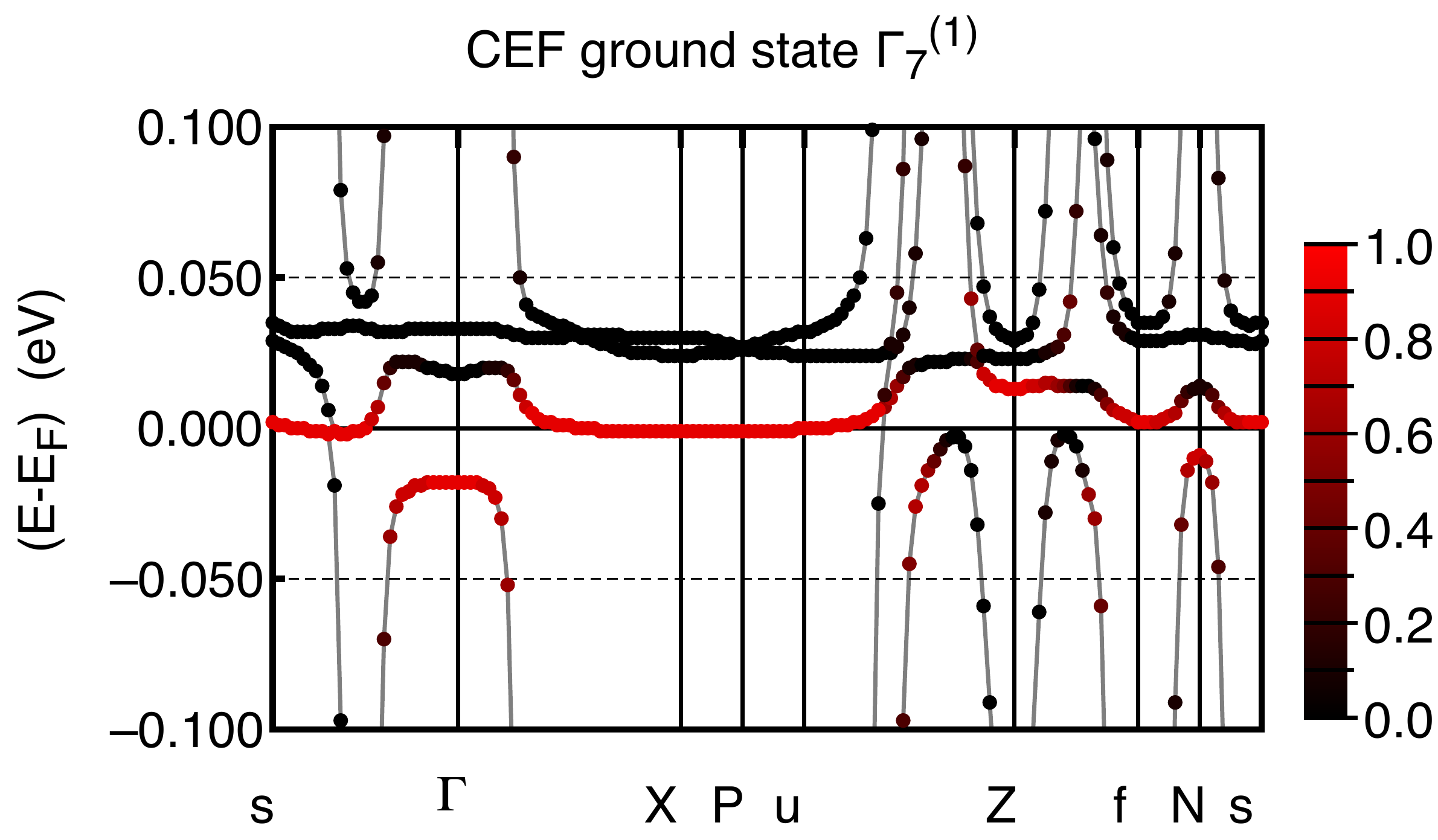}
		\includegraphics[width=0.9\columnwidth]{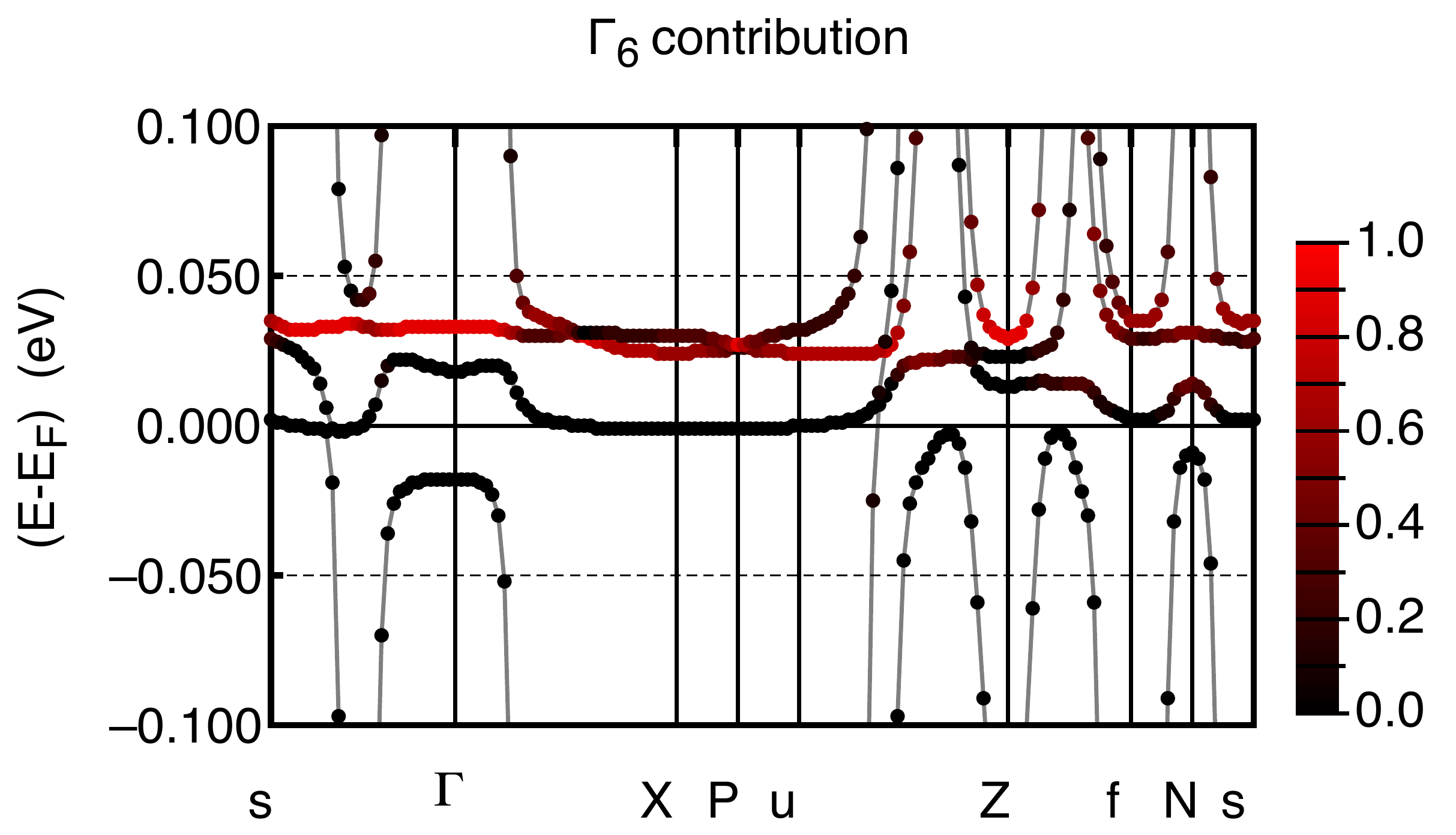}
		\includegraphics[width=0.9\columnwidth]{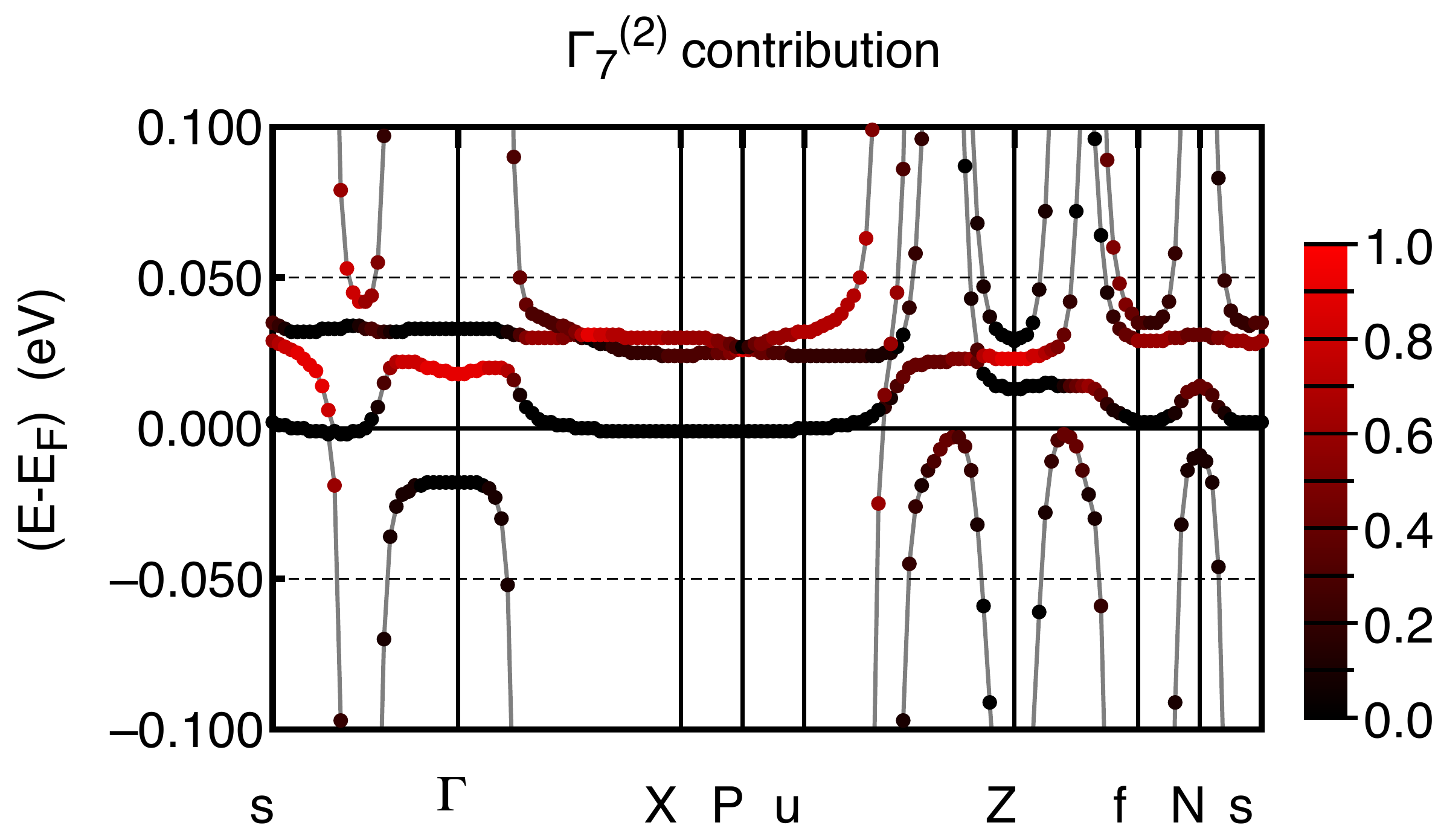}
    \caption{(color online) Renormalized quasiparticle bands, showing the contributions of each crystal-field state.}
    \label{Zwicknagl}
\end{figure}

The small but non-negligible amount of hybridization-induced mixing-in of the excited states into the ground state is also very much supported by the results of re-normalized band-structure calculations \cite{Zwicknagl1992,Zwick1993}. Here the present ground-state symmetry and excited 4$f$ states at 30\,meV are taken into account and the contributions of the three crystal-field states are projected out to the respective bands (see Fig.\,\ref{Zwicknagl}). The Fermi surface has two major sheets with quasiparticle masses that differ profoundly in which the Fermi surface sheet with the light quasiparticles very much resembles that of the LDA standard band structure calculations. It indeed turns out that the heavy band at zero energy has mainly $\Gamma_7^{(1)}$ character with some \sout{minor} contributions in the few percent range of the $\Gamma_7^{(2)}$ state, and, due to a smaller hybridization function, somewhat less of the $\Gamma_6$. 

Finally, for completeness, we need to examine the consequences of having a mixed-orbital ground state for the analysis of the LD in XAS. We calculated the XAS spectra for a ground state which consists of 88\% $\Gamma_7^{(1)}$, 6\% of $\Gamma_6$, and 6\% of $\Gamma_7^{(2)}$ states. This requires the adjustment of the $\alpha$ value to $\left|0.62\right|$ in order to still reproduce the experiment. The correction is minor not only because the contribution of the higher lying states is small, in addition, the LD of the $\Gamma_6$ and $\Gamma_7^{(2)}$ states partially cancel each other out. The small correction to $\alpha$ makes us confident that the crystal-field model with the Boltzmann occupation of the excited states is of more than sufficient accuracy to support our conclusion that no orbital switching as function of temperature takes place. Instead, we can safely infer the presence of a robust static crystal-field scheme although, in principle, the analysis of the temperature dependence of the LD would require a temperature dependent Anderson impurity calculation. It would also be very interesting to study quantitatively in the same frame work the impact of the mixing-in of the $\Gamma_6$ and $\Gamma_7^{(2)}$ on the magnetic susceptibility.

The present data, the Anderson impurity as well as LDA calculations principally allow for a multiorbital ground state that includes the $\Gamma_{6}$ crystal-field state that is expected for the $d$+$d$ singlet pairing state as proposed by Nica and Si. Their analysis involves $\Gamma_{6}$ Wannier orbitals of the conduction electron states near the Fermi energy. These Wannier orbitals will hybridize with the excited $\Gamma_{6}$ crystal-field state of the $f$ manifold and thereby make it a small but nonzero component in the ground-state\,\cite{Emilian2019,Priv2020}.

\section{Summary}
CeCu$_2$Si$_2$ has been investigated with soft x-ray absorption spectroscopy in the temperature range from 250\,mK to 250\,K and the $J_z$ admixture of the $\Gamma_7$ ground state wave function has been determined. The overall temperature dependence of the experimental linear dichroism LD($T$) is well reproduced by the thermal occupation of excited crystal-field states so that the scenario of orbital switching seems unlikely. The spectra indicate the presence of the Ce 4$f^0$ configuration in the ground state so that in principle the ground state can have a multi-orbital character. Based on the experimentally confirmed Kondo temperature and the crystal-field energies, the contribution of the higher lying $\Gamma_6$ and $\Gamma_7^{(2)}$ crystal-field states to the primarily $\Gamma_7^{(1)}$ ground state is estimated to be about 6\% within the 4$f^1$ manifold. This estimate is supported by re-normalized band structure calculations that uses the experimentally determined crystal-field scheme. 
 
\section{Acknowledgment} 
We would like to thank P. Thalmeier for various discussion and input. We further acknowledge Synchrotron-SOLEIL and synchrotron-ALBA (exp. ID  2019023540) for provision of synchrotron radiation facilities and the support from the Max Planck-POSTECH-Hsinchu Center for Complex Phase Materials. This work has benefited from a funding from LabEx PALM (ANR-10-LABX-0039-PALM) and M.S. and A.S. gratefully acknowledge the financial support of the German Funding Agency Deutsche Forschungsgemeinschaft through grants SE\,1441-4-1, and SE\,1441-5-1. G. Z. acknowledges financial support through the ANR-DFG project Fermi-NESt (Deutsche Forschungsgemeinschaft Grant No ZW 77/5-1).


\begin{thebibliography}{51}%
\makeatletter
\providecommand \@ifxundefined [1]{%
 \@ifx{#1\undefined}
}%
\providecommand \@ifnum [1]{%
 \ifnum #1\expandafter \@firstoftwo
 \else \expandafter \@secondoftwo
 \fi
}%
\providecommand \@ifx [1]{%
 \ifx #1\expandafter \@firstoftwo
 \else \expandafter \@secondoftwo
 \fi
}%
\providecommand \natexlab [1]{#1}%
\providecommand \enquote  [1]{``#1''}%
\providecommand \bibnamefont  [1]{#1}%
\providecommand \bibfnamefont [1]{#1}%
\providecommand \citenamefont [1]{#1}%
\providecommand \href@noop [0]{\@secondoftwo}%
\providecommand \href [0]{\begingroup \@sanitize@url \@href}%
\providecommand \@href[1]{\@@startlink{#1}\@@href}%
\providecommand \@@href[1]{\endgroup#1\@@endlink}%
\providecommand \@sanitize@url [0]{\catcode `\\12\catcode `\$12\catcode
  `\&12\catcode `\#12\catcode `\^12\catcode `\_12\catcode `\%12\relax}%
\providecommand \@@startlink[1]{}%
\providecommand \@@endlink[0]{}%
\providecommand \url  [0]{\begingroup\@sanitize@url \@url }%
\providecommand \@url [1]{\endgroup\@href {#1}{\urlprefix }}%
\providecommand \urlprefix  [0]{URL }%
\providecommand \Eprint [0]{\href }%
\providecommand \doibase [0]{https://doi.org/}%
\providecommand \selectlanguage [0]{\@gobble}%
\providecommand \bibinfo  [0]{\@secondoftwo}%
\providecommand \bibfield  [0]{\@secondoftwo}%
\providecommand \translation [1]{[#1]}%
\providecommand \BibitemOpen [0]{}%
\providecommand \bibitemStop [0]{}%
\providecommand \bibitemNoStop [0]{.\EOS\space}%
\providecommand \EOS [0]{\spacefactor3000\relax}%
\providecommand \BibitemShut  [1]{\csname bibitem#1\endcsname}%
\let\auto@bib@innerbib\@empty
%</preamble>
\bibitem [{\citenamefont {Coleman}(2007)}]{Coleman2007}%
  \BibitemOpen
  \bibfield  {author} {\bibinfo {author} {\bibfnamefont {P.}~\bibnamefont
  {Coleman}},\ }\href@noop {} {\emph {\bibinfo {title} {{H}andbook of {M}agn.
  and {A}dv. {M}agn. {M}ater.}}},\ edited by\ \bibinfo {editor} {\bibfnamefont
  {M.~F. S.~M.}\ \bibnamefont {H.~Kronm\"uller}, \bibfnamefont {S.~Parkin}}\
  and\ \bibinfo {editor} {\bibfnamefont {I.}~\bibnamefont {Zutic}},\
  Vol.~\bibinfo {volume} {1}\ (\bibinfo  {publisher} {John Wiley and Sons},\
  \bibinfo {year} {2007})\ pp.\ \bibinfo {pages} {95--148},\ \bibinfo {note}
  {"Heavy fermions: electrons at the edge of magnetism"}\BibitemShut {NoStop}%
\bibitem [{\citenamefont {Coleman}(2015)}]{Coleman2015}%
  \BibitemOpen
  \bibfield  {author} {\bibinfo {author} {\bibfnamefont {P.}~\bibnamefont
  {Coleman}},\ }\bibfield  {title} {\bibinfo {title} {{Heavy fermions and the
  Kondo lattice: A 21st century perspective}},\ }\href@noop {} {\bibfield
  {journal} {\bibinfo  {journal} {edt. E. Pavarini, E. Koch, and P. Coleman}\
  }\textbf {\bibinfo {volume} {(Forschungszentrum J\"ulich)}},\ \bibinfo
  {pages} {95} (\bibinfo {year} {2015})},\ \bibinfo {note} {{in Many-Body
  Physics: From Kondo to Hubbard}}\BibitemShut {NoStop}%
\bibitem [{\citenamefont {L\"ohneysen}\ \emph {et~al.}(2007)\citenamefont
  {L\"ohneysen}, \citenamefont {Rosch}, \citenamefont {Vojta},\ and\
  \citenamefont {W\"olfle}}]{Hilbert2007}%
  \BibitemOpen
  \bibfield  {author} {\bibinfo {author} {\bibfnamefont {H.~v.}\ \bibnamefont
  {L\"ohneysen}}, \bibinfo {author} {\bibfnamefont {A.}~\bibnamefont {Rosch}},
  \bibinfo {author} {\bibfnamefont {M.}~\bibnamefont {Vojta}},\ and\ \bibinfo
  {author} {\bibfnamefont {P.}~\bibnamefont {W\"olfle}},\ }\bibfield  {title}
  {\bibinfo {title} {Fermi-liquid instabilities at magnetic quantum phase
  transitions},\ }\href {https://doi.org/10.1103/RevModPhys.79.1015} {\bibfield
   {journal} {\bibinfo  {journal} {Rev. Mod. Phys.}\ }\textbf {\bibinfo
  {volume} {79}},\ \bibinfo {pages} {1015} (\bibinfo {year}
  {2007})}\BibitemShut {NoStop}%
\bibitem [{\citenamefont {Wirth}\ and\ \citenamefont
  {Steglich}(2016)}]{Wirth2016}%
  \BibitemOpen
  \bibfield  {author} {\bibinfo {author} {\bibfnamefont {S.}~\bibnamefont
  {Wirth}}\ and\ \bibinfo {author} {\bibfnamefont {F.}~\bibnamefont
  {Steglich}},\ }\bibfield  {title} {\bibinfo {title} {Exploring heavy fermions
  from macroscopic to microscopic length scales},\ }\href
  {https://doi.org/10.1038/natrevmats.2016.51} {\bibfield  {journal} {\bibinfo
  {journal} {Nat. Rev. Mater.}\ }\textbf {\bibinfo {volume} {1}},\ \bibinfo
  {pages} {16066} (\bibinfo {year} {2016})}\BibitemShut {NoStop}%
\bibitem [{\citenamefont {Sun}\ and\ \citenamefont {Steglich}(2013)}]{Sun2013}%
  \BibitemOpen
  \bibfield  {author} {\bibinfo {author} {\bibfnamefont {P.}~\bibnamefont
  {Sun}}\ and\ \bibinfo {author} {\bibfnamefont {F.}~\bibnamefont {Steglich}},\
  }\bibfield  {title} {\bibinfo {title} {Nernst effect: Evidence of local
  {K}ondo scattering in heavy fermions},\ }\href
  {https://doi.org/10.1103/PhysRevLett.110.216408} {\bibfield  {journal}
  {\bibinfo  {journal} {Phys. Rev. Lett.}\ }\textbf {\bibinfo {volume} {110}},\
  \bibinfo {pages} {216408} (\bibinfo {year} {2013})}\BibitemShut {NoStop}%
\bibitem [{\citenamefont {Steglich}\ and\ \citenamefont
  {Wirth}(2016)}]{Steglich2016}%
  \BibitemOpen
  \bibfield  {author} {\bibinfo {author} {\bibfnamefont {F.}~\bibnamefont
  {Steglich}}\ and\ \bibinfo {author} {\bibfnamefont {S.}~\bibnamefont
  {Wirth}},\ }\bibfield  {title} {\bibinfo {title} {Foundations of
  heavy-fermion superconductivity: lattice {K}ondo effect and {M}ott physics},\
  }\href {https://doi.org/10.1088/0034-4885/79/8/084502} {\bibfield  {journal}
  {\bibinfo  {journal} {Rep. Prof. Phys.}\ }\textbf {\bibinfo {volume} {79}},\
  \bibinfo {pages} {084502} (\bibinfo {year} {2016})}\BibitemShut {NoStop}%
\bibitem [{\citenamefont {Steglich}\ \emph {et~al.}(1979)\citenamefont
  {Steglich}, \citenamefont {Aarts}, \citenamefont {Bredl}, \citenamefont
  {Lieke}, \citenamefont {Meschede}, \citenamefont {Franz},\ and\ \citenamefont
  {Sch\"afer}}]{Steglich1979}%
  \BibitemOpen
  \bibfield  {author} {\bibinfo {author} {\bibfnamefont {F.}~\bibnamefont
  {Steglich}}, \bibinfo {author} {\bibfnamefont {J.}~\bibnamefont {Aarts}},
  \bibinfo {author} {\bibfnamefont {C.~D.}\ \bibnamefont {Bredl}}, \bibinfo
  {author} {\bibfnamefont {W.}~\bibnamefont {Lieke}}, \bibinfo {author}
  {\bibfnamefont {D.}~\bibnamefont {Meschede}}, \bibinfo {author}
  {\bibfnamefont {W.}~\bibnamefont {Franz}},\ and\ \bibinfo {author}
  {\bibfnamefont {H.}~\bibnamefont {Sch\"afer}},\ }\bibfield  {title} {\bibinfo
  {title} {Superconductivity in the presence of strong {P}auli paramagnetism:
  {CeC}u$_2${S}i$_2$},\ }\href {https://doi.org/10.1103/PhysRevLett.43.1892}
  {\bibfield  {journal} {\bibinfo  {journal} {Phys. Rev. Lett.}\ }\textbf
  {\bibinfo {volume} {43}},\ \bibinfo {pages} {1892} (\bibinfo {year}
  {1979})}\BibitemShut {NoStop}%
\bibitem [{\citenamefont {Thomas}\ \emph {et~al.}(1993)\citenamefont {Thomas},
  \citenamefont {Thomasson}, \citenamefont {Ayache}, \citenamefont {Geibel},\
  and\ \citenamefont {Steglich}}]{Thomas1993}%
  \BibitemOpen
  \bibfield  {author} {\bibinfo {author} {\bibfnamefont {F.}~\bibnamefont
  {Thomas}}, \bibinfo {author} {\bibfnamefont {J.}~\bibnamefont {Thomasson}},
  \bibinfo {author} {\bibfnamefont {C.}~\bibnamefont {Ayache}}, \bibinfo
  {author} {\bibfnamefont {C.}~\bibnamefont {Geibel}},\ and\ \bibinfo {author}
  {\bibfnamefont {F.}~\bibnamefont {Steglich}},\ }\bibfield  {title} {\bibinfo
  {title} {Precise determination of the pressure dependence of {T}$_c$ in the
  heavy-fermion superconductor {CeCu$_2$Si$_2$}},\ }\href@noop {} {\bibfield
  {journal} {\bibinfo  {journal} {Physica B}\ }\textbf {\bibinfo {volume}
  {186-188}},\ \bibinfo {pages} {303} (\bibinfo {year} {1993})}\BibitemShut
  {NoStop}%
\bibitem [{\citenamefont {Yuan}\ \emph {et~al.}(2003)\citenamefont {Yuan},
  \citenamefont {Grosche}, \citenamefont {Deppe}, \citenamefont {Geibel},
  \citenamefont {Sparn},\ and\ \citenamefont {Steglich}}]{Yuan2003}%
  \BibitemOpen
  \bibfield  {author} {\bibinfo {author} {\bibfnamefont {H.~Q.}\ \bibnamefont
  {Yuan}}, \bibinfo {author} {\bibfnamefont {F.~M.}\ \bibnamefont {Grosche}},
  \bibinfo {author} {\bibfnamefont {M.}~\bibnamefont {Deppe}}, \bibinfo
  {author} {\bibfnamefont {C.}~\bibnamefont {Geibel}}, \bibinfo {author}
  {\bibfnamefont {G.}~\bibnamefont {Sparn}},\ and\ \bibinfo {author}
  {\bibfnamefont {F.}~\bibnamefont {Steglich}},\ }\bibfield  {title} {\bibinfo
  {title} {Observation of two distinct superconducting phases in
  {CeCu$_2$S}i$_2$},\ }\href {https://doi.org/10.1126/science.1091648}
  {\bibfield  {journal} {\bibinfo  {journal} {Science}\ }\textbf {\bibinfo
  {volume} {302}},\ \bibinfo {pages} {2104} (\bibinfo {year}
  {2003})}\BibitemShut {NoStop}%
\bibitem [{\citenamefont {Stockert}\ \emph {et~al.}(2010)\citenamefont
  {Stockert}, \citenamefont {Arndt}, \citenamefont {Faulhaber}, \citenamefont
  {Geibel}, \citenamefont {Jeevan}, \citenamefont {Kirchner}, \citenamefont
  {Loewenhaupt}, \citenamefont {Schmalzl}, \citenamefont {Schmidt},
  \citenamefont {Si},\ and\ \citenamefont {Steglich}}]{Stockert2010}%
  \BibitemOpen
  \bibfield  {author} {\bibinfo {author} {\bibfnamefont {O.}~\bibnamefont
  {Stockert}}, \bibinfo {author} {\bibfnamefont {J.}~\bibnamefont {Arndt}},
  \bibinfo {author} {\bibfnamefont {E.}~\bibnamefont {Faulhaber}}, \bibinfo
  {author} {\bibfnamefont {C.}~\bibnamefont {Geibel}}, \bibinfo {author}
  {\bibfnamefont {H.~S.}\ \bibnamefont {Jeevan}}, \bibinfo {author}
  {\bibfnamefont {S.}~\bibnamefont {Kirchner}}, \bibinfo {author}
  {\bibfnamefont {M.}~\bibnamefont {Loewenhaupt}}, \bibinfo {author}
  {\bibfnamefont {K.}~\bibnamefont {Schmalzl}}, \bibinfo {author}
  {\bibfnamefont {W.}~\bibnamefont {Schmidt}}, \bibinfo {author} {\bibfnamefont
  {Q.}~\bibnamefont {Si}},\ and\ \bibinfo {author} {\bibfnamefont
  {F.}~\bibnamefont {Steglich}},\ }\bibfield  {title} {\bibinfo {title}
  {Magnetically driven superconductivity in {CeCu$_2$S}i$_2$},\ }\href
  {https://doi.org/10.1038/nphys1852} {\bibfield  {journal} {\bibinfo
  {journal} {Nat. Phys.}\ }\textbf {\bibinfo {volume} {7}},\ \bibinfo {pages}
  {119} (\bibinfo {year} {2010})}\BibitemShut {NoStop}%
\bibitem [{\citenamefont {Eremin}\ \emph {et~al.}(2008)\citenamefont {Eremin},
  \citenamefont {Zwicknagl}, \citenamefont {Thalmeier},\ and\ \citenamefont
  {Fulde}}]{Eremin2008}%
  \BibitemOpen
  \bibfield  {author} {\bibinfo {author} {\bibfnamefont {I.}~\bibnamefont
  {Eremin}}, \bibinfo {author} {\bibfnamefont {G.}~\bibnamefont {Zwicknagl}},
  \bibinfo {author} {\bibfnamefont {P.}~\bibnamefont {Thalmeier}},\ and\
  \bibinfo {author} {\bibfnamefont {P.}~\bibnamefont {Fulde}},\ }\bibfield
  {title} {\bibinfo {title} {Feedback spin resonance in superconducting
  {CeCu}$_2${Si}$_2$ and {CeCoIn}$_5$},\ }\href
  {https://doi.org/10.1103/PhysRevLett.101.187001} {\bibfield  {journal}
  {\bibinfo  {journal} {Phys. Rev. Lett.}\ }\textbf {\bibinfo {volume} {101}},\
  \bibinfo {pages} {187001} (\bibinfo {year} {2008})}\BibitemShut {NoStop}%
\bibitem [{\citenamefont {Stockert}\ and\ \citenamefont
  {Steglich}(2011)}]{Stockert2011}%
  \BibitemOpen
  \bibfield  {author} {\bibinfo {author} {\bibfnamefont {O.}~\bibnamefont
  {Stockert}}\ and\ \bibinfo {author} {\bibfnamefont {F.}~\bibnamefont
  {Steglich}},\ }\bibfield  {title} {\bibinfo {title} {Unconventional quantum
  criticality in heavy-fermion compounds},\ }\href
  {https://doi.org/10.1146/annurev-conmatphys-062910-140546} {\bibfield
  {journal} {\bibinfo  {journal} {Annual Review of Condensed Matter Physics}\
  }\textbf {\bibinfo {volume} {2}},\ \bibinfo {pages} {79} (\bibinfo {year}
  {2011})}\BibitemShut {NoStop}%
\bibitem [{\citenamefont {Pang}\ \emph {et~al.}(2018)\citenamefont {Pang},
  \citenamefont {Smidman}, \citenamefont {Zhang}, \citenamefont {Jiao},
  \citenamefont {Weng}, \citenamefont {Nica}, \citenamefont {Chen},
  \citenamefont {Jiang}, \citenamefont {Zhang}, \citenamefont {Xie},
  \citenamefont {Jeevan}, \citenamefont {Lee}, \citenamefont {Gegenwart},
  \citenamefont {Steglich}, \citenamefont {Si},\ and\ \citenamefont
  {Yuan}}]{Pang2018}%
  \BibitemOpen
  \bibfield  {author} {\bibinfo {author} {\bibfnamefont {G.}~\bibnamefont
  {Pang}}, \bibinfo {author} {\bibfnamefont {M.}~\bibnamefont {Smidman}},
  \bibinfo {author} {\bibfnamefont {J.}~\bibnamefont {Zhang}}, \bibinfo
  {author} {\bibfnamefont {L.}~\bibnamefont {Jiao}}, \bibinfo {author}
  {\bibfnamefont {Z.}~\bibnamefont {Weng}}, \bibinfo {author} {\bibfnamefont
  {E.~M.}\ \bibnamefont {Nica}}, \bibinfo {author} {\bibfnamefont
  {Y.}~\bibnamefont {Chen}}, \bibinfo {author} {\bibfnamefont {W.}~\bibnamefont
  {Jiang}}, \bibinfo {author} {\bibfnamefont {Y.}~\bibnamefont {Zhang}},
  \bibinfo {author} {\bibfnamefont {W.}~\bibnamefont {Xie}}, \bibinfo {author}
  {\bibfnamefont {H.~S.}\ \bibnamefont {Jeevan}}, \bibinfo {author}
  {\bibfnamefont {H.}~\bibnamefont {Lee}}, \bibinfo {author} {\bibfnamefont
  {P.}~\bibnamefont {Gegenwart}}, \bibinfo {author} {\bibfnamefont
  {F.}~\bibnamefont {Steglich}}, \bibinfo {author} {\bibfnamefont
  {Q.}~\bibnamefont {Si}},\ and\ \bibinfo {author} {\bibfnamefont
  {H.}~\bibnamefont {Yuan}},\ }\bibfield  {title} {\bibinfo {title} {Fully
  gapped $d$-wave superconductivity in {CeC}u$_2${S}i$_2$},\ }\href
  {https://doi.org/10.1073/pnas.1720291115} {\bibfield  {journal} {\bibinfo
  {journal} {Proc. Nat. Acad. Science. U.S.A.}\ }\textbf {\bibinfo {volume}
  {115}},\ \bibinfo {pages} {5343} (\bibinfo {year} {2018})}\BibitemShut
  {NoStop}%
\bibitem [{\citenamefont {Holmes}\ \emph {et~al.}(2007)\citenamefont {Holmes},
  \citenamefont {Jaccard},\ and\ \citenamefont {Miyake}}]{Holmes2007}%
  \BibitemOpen
  \bibfield  {author} {\bibinfo {author} {\bibfnamefont {A.~T.}\ \bibnamefont
  {Holmes}}, \bibinfo {author} {\bibfnamefont {D.}~\bibnamefont {Jaccard}},\
  and\ \bibinfo {author} {\bibfnamefont {K.}~\bibnamefont {Miyake}},\
  }\bibfield  {title} {\bibinfo {title} {Valence instability and
  superconductivity in heavy fermion systems},\ }\href
  {https://doi.org/10.1143/JPSJ.76.051002} {\bibfield  {journal} {\bibinfo
  {journal} {J. Phys. Soc. Jpn.}\ }\textbf {\bibinfo {volume} {76}},\ \bibinfo
  {pages} {051002} (\bibinfo {year} {2007})}\BibitemShut {NoStop}%
\bibitem [{\citenamefont {Rueff}\ \emph {et~al.}(2011)\citenamefont {Rueff},
  \citenamefont {Raymond}, \citenamefont {Taguchi}, \citenamefont {Sikora},
  \citenamefont {Iti\'e}, \citenamefont {Baudelet}, \citenamefont
  {Braithwaite}, \citenamefont {Knebel},\ and\ \citenamefont
  {Jaccard}}]{Rueff2011}%
  \BibitemOpen
  \bibfield  {author} {\bibinfo {author} {\bibfnamefont {J.-P.}\ \bibnamefont
  {Rueff}}, \bibinfo {author} {\bibfnamefont {S.}~\bibnamefont {Raymond}},
  \bibinfo {author} {\bibfnamefont {M.}~\bibnamefont {Taguchi}}, \bibinfo
  {author} {\bibfnamefont {M.}~\bibnamefont {Sikora}}, \bibinfo {author}
  {\bibfnamefont {J.-P.}\ \bibnamefont {Iti\'e}}, \bibinfo {author}
  {\bibfnamefont {F.}~\bibnamefont {Baudelet}}, \bibinfo {author}
  {\bibfnamefont {D.}~\bibnamefont {Braithwaite}}, \bibinfo {author}
  {\bibfnamefont {G.}~\bibnamefont {Knebel}},\ and\ \bibinfo {author}
  {\bibfnamefont {D.}~\bibnamefont {Jaccard}},\ }\bibfield  {title} {\bibinfo
  {title} {Pressure-induced valence crossover in superconducting
  {CeCu}$_2${Si}$_2$},\ }\href {https://doi.org/10.1103/PhysRevLett.106.186405}
  {\bibfield  {journal} {\bibinfo  {journal} {Phys. Rev. Lett.}\ }\textbf
  {\bibinfo {volume} {106}},\ \bibinfo {pages} {186405} (\bibinfo {year}
  {2011})}\BibitemShut {NoStop}%
\bibitem [{\citenamefont {Kittaka}\ \emph {et~al.}(2014)\citenamefont
  {Kittaka}, \citenamefont {Aoki}, \citenamefont {Shimura}, \citenamefont
  {Sakakibara}, \citenamefont {Seiro}, \citenamefont {Geibel}, \citenamefont
  {Steglich}, \citenamefont {Ikeda},\ and\ \citenamefont
  {Machida}}]{Kittaka2014}%
  \BibitemOpen
  \bibfield  {author} {\bibinfo {author} {\bibfnamefont {S.}~\bibnamefont
  {Kittaka}}, \bibinfo {author} {\bibfnamefont {Y.}~\bibnamefont {Aoki}},
  \bibinfo {author} {\bibfnamefont {Y.}~\bibnamefont {Shimura}}, \bibinfo
  {author} {\bibfnamefont {T.}~\bibnamefont {Sakakibara}}, \bibinfo {author}
  {\bibfnamefont {S.}~\bibnamefont {Seiro}}, \bibinfo {author} {\bibfnamefont
  {C.}~\bibnamefont {Geibel}}, \bibinfo {author} {\bibfnamefont
  {F.}~\bibnamefont {Steglich}}, \bibinfo {author} {\bibfnamefont
  {H.}~\bibnamefont {Ikeda}},\ and\ \bibinfo {author} {\bibfnamefont
  {K.}~\bibnamefont {Machida}},\ }\bibfield  {title} {\bibinfo {title}
  {Multiband superconductivity with unexpected deficiency of nodal
  quasiparticles in {CeCu}$_2${Si}$_2$},\ }\href
  {https://doi.org/10.1103/PhysRevLett.112.067002} {\bibfield  {journal}
  {\bibinfo  {journal} {Phys. Rev. Lett.}\ }\textbf {\bibinfo {volume} {112}},\
  \bibinfo {pages} {067002} (\bibinfo {year} {2014})}\BibitemShut {NoStop}%
\bibitem [{\citenamefont {Yamashita}\ \emph {et~al.}(2017)\citenamefont
  {Yamashita}, \citenamefont {Takenaka}, \citenamefont {Tokiwa}, \citenamefont
  {Wilcox}, \citenamefont {Mizukami}, \citenamefont {Terazawa}, \citenamefont
  {Kasahara}, \citenamefont {Kittaka}, \citenamefont {Sakakibara},
  \citenamefont {Konczykowski}, \citenamefont {Seiro}, \citenamefont {Jeevan},
  \citenamefont {Geibel}, \citenamefont {Putzke}, \citenamefont {Onishi},
  \citenamefont {Ikeda}, \citenamefont {Carrington}, \citenamefont
  {Shibauchi},\ and\ \citenamefont {Matsuda}}]{Yamashitae2017}%
  \BibitemOpen
  \bibfield  {author} {\bibinfo {author} {\bibfnamefont {T.}~\bibnamefont
  {Yamashita}}, \bibinfo {author} {\bibfnamefont {T.}~\bibnamefont {Takenaka}},
  \bibinfo {author} {\bibfnamefont {Y.}~\bibnamefont {Tokiwa}}, \bibinfo
  {author} {\bibfnamefont {J.~A.}\ \bibnamefont {Wilcox}}, \bibinfo {author}
  {\bibfnamefont {Y.}~\bibnamefont {Mizukami}}, \bibinfo {author}
  {\bibfnamefont {D.}~\bibnamefont {Terazawa}}, \bibinfo {author}
  {\bibfnamefont {Y.}~\bibnamefont {Kasahara}}, \bibinfo {author}
  {\bibfnamefont {S.}~\bibnamefont {Kittaka}}, \bibinfo {author} {\bibfnamefont
  {T.}~\bibnamefont {Sakakibara}}, \bibinfo {author} {\bibfnamefont
  {M.}~\bibnamefont {Konczykowski}}, \bibinfo {author} {\bibfnamefont
  {S.}~\bibnamefont {Seiro}}, \bibinfo {author} {\bibfnamefont {H.~S.}\
  \bibnamefont {Jeevan}}, \bibinfo {author} {\bibfnamefont {C.}~\bibnamefont
  {Geibel}}, \bibinfo {author} {\bibfnamefont {C.}~\bibnamefont {Putzke}},
  \bibinfo {author} {\bibfnamefont {T.}~\bibnamefont {Onishi}}, \bibinfo
  {author} {\bibfnamefont {H.}~\bibnamefont {Ikeda}}, \bibinfo {author}
  {\bibfnamefont {A.}~\bibnamefont {Carrington}}, \bibinfo {author}
  {\bibfnamefont {T.}~\bibnamefont {Shibauchi}},\ and\ \bibinfo {author}
  {\bibfnamefont {Y.}~\bibnamefont {Matsuda}},\ }\bibfield  {title} {\bibinfo
  {title} {Fully gapped superconductivity with no sign change in the
  prototypical heavy-fermion {CeCu}$_2${Si}$_2$},\ }\bibfield  {journal}
  {\bibinfo  {journal} {Sci. Adv.}\ }\textbf {\bibinfo {volume} {3}},\ \href
  {https://doi.org/10.1126/sciadv.1601667} {10.1126/sciadv.1601667} (\bibinfo
  {year} {2017})\BibitemShut {NoStop}%
\bibitem [{\citenamefont {Takenaka}\ \emph {et~al.}(2017)\citenamefont
  {Takenaka}, \citenamefont {Mizukami}, \citenamefont {Wilcox}, \citenamefont
  {Konczykowski}, \citenamefont {Seiro}, \citenamefont {Geibel}, \citenamefont
  {Tokiwa}, \citenamefont {Kasahara}, \citenamefont {Putzke}, \citenamefont
  {Matsuda}, \citenamefont {Carrington},\ and\ \citenamefont
  {Shibauchi}}]{Takenaka2017}%
  \BibitemOpen
  \bibfield  {author} {\bibinfo {author} {\bibfnamefont {T.}~\bibnamefont
  {Takenaka}}, \bibinfo {author} {\bibfnamefont {Y.}~\bibnamefont {Mizukami}},
  \bibinfo {author} {\bibfnamefont {J.~A.}\ \bibnamefont {Wilcox}}, \bibinfo
  {author} {\bibfnamefont {M.}~\bibnamefont {Konczykowski}}, \bibinfo {author}
  {\bibfnamefont {S.}~\bibnamefont {Seiro}}, \bibinfo {author} {\bibfnamefont
  {C.}~\bibnamefont {Geibel}}, \bibinfo {author} {\bibfnamefont
  {Y.}~\bibnamefont {Tokiwa}}, \bibinfo {author} {\bibfnamefont
  {Y.}~\bibnamefont {Kasahara}}, \bibinfo {author} {\bibfnamefont
  {C.}~\bibnamefont {Putzke}}, \bibinfo {author} {\bibfnamefont
  {Y.}~\bibnamefont {Matsuda}}, \bibinfo {author} {\bibfnamefont
  {A.}~\bibnamefont {Carrington}},\ and\ \bibinfo {author} {\bibfnamefont
  {T.}~\bibnamefont {Shibauchi}},\ }\bibfield  {title} {\bibinfo {title}
  {Full-gap superconductivity robust against disorder in heavy-fermion
  {CeCu}$_2${Si}$_2$},\ }\href {https://doi.org/10.1103/PhysRevLett.119.077001}
  {\bibfield  {journal} {\bibinfo  {journal} {Phys. Rev. Lett.}\ }\textbf
  {\bibinfo {volume} {119}},\ \bibinfo {pages} {077001} (\bibinfo {year}
  {2017})}\BibitemShut {NoStop}%
\bibitem [{\citenamefont {Nica}\ and\ \citenamefont {Si}(2019)}]{Emilian2019}%
  \BibitemOpen
  \bibfield  {author} {\bibinfo {author} {\bibfnamefont {E.~M.}\ \bibnamefont
  {Nica}}\ and\ \bibinfo {author} {\bibfnamefont {Q.}~\bibnamefont {Si}},\
  }\href@noop {} {\bibinfo {title} {Multiorbital singlet pairing and $d+d$
  superconductivity}} (\bibinfo {year} {2019}),\ \Eprint
  {https://arxiv.org/abs/1911.13274} {arXiv:1911.13274 [cond-mat.supr-con]}
  \BibitemShut {NoStop}%
\bibitem [{\citenamefont {Pourovskii}\ \emph {et~al.}(2014)\citenamefont
  {Pourovskii}, \citenamefont {Hansmann}, \citenamefont {Ferrero},\ and\
  \citenamefont {Georges}}]{Pourovskii2014}%
  \BibitemOpen
  \bibfield  {author} {\bibinfo {author} {\bibfnamefont {L.~V.}\ \bibnamefont
  {Pourovskii}}, \bibinfo {author} {\bibfnamefont {P.}~\bibnamefont
  {Hansmann}}, \bibinfo {author} {\bibfnamefont {M.}~\bibnamefont {Ferrero}},\
  and\ \bibinfo {author} {\bibfnamefont {A.}~\bibnamefont {Georges}},\
  }\bibfield  {title} {\bibinfo {title} {Theoretical prediction and
  spectroscopic fingerprints of an orbital transition in {CeCu}$_2${Si}$_2$},\
  }\href {https://doi.org/10.1103/PhysRevLett.112.106407} {\bibfield  {journal}
  {\bibinfo  {journal} {Phys. Rev. Lett.}\ }\textbf {\bibinfo {volume} {112}},\
  \bibinfo {pages} {106407} (\bibinfo {year} {2014})}\BibitemShut {NoStop}%
\bibitem [{\citenamefont {Goremychkin}\ and\ \citenamefont
  {Osborn}(1993)}]{Goremychkin1993}%
  \BibitemOpen
  \bibfield  {author} {\bibinfo {author} {\bibfnamefont {E.~A.}\ \bibnamefont
  {Goremychkin}}\ and\ \bibinfo {author} {\bibfnamefont {R.}~\bibnamefont
  {Osborn}},\ }\bibfield  {title} {\bibinfo {title} {Crystal-field excitations
  in {CeCu}$_2${Si}$_2$},\ }\href {https://doi.org/10.1103/PhysRevB.47.14280}
  {\bibfield  {journal} {\bibinfo  {journal} {Phys. Rev. B}\ }\textbf {\bibinfo
  {volume} {47}},\ \bibinfo {pages} {14280} (\bibinfo {year} {1993})},\
  \bibinfo {note} {note:\,on page 14287 $\eta$ and $\sqrt{(1-\eta^2)}$ are
  exchanged. Correct is $|g.s.\rangle$\,=\,$\sqrt{(1-\eta^2)}$
  $|$$\pm$5/2$\rangle$ $\pm$ $\eta$$|$$\mp$3/2$\rangle$ with
  $\sqrt{(1-\eta^2)}$=0.88 to reproduce the anisotropy of the static
  susceptibility.}\BibitemShut {Stop}%
\bibitem [{\citenamefont {Haverkort}\ \emph {et~al.}(2007)\citenamefont
  {Haverkort}, \citenamefont {Tanaka}, \citenamefont {Tjeng},\ and\
  \citenamefont {Sawatzky}}]{Haverkort2007}%
  \BibitemOpen
  \bibfield  {author} {\bibinfo {author} {\bibfnamefont {M.~W.}\ \bibnamefont
  {Haverkort}}, \bibinfo {author} {\bibfnamefont {A.}~\bibnamefont {Tanaka}},
  \bibinfo {author} {\bibfnamefont {L.~H.}\ \bibnamefont {Tjeng}},\ and\
  \bibinfo {author} {\bibfnamefont {G.~A.}\ \bibnamefont {Sawatzky}},\
  }\bibfield  {title} {\bibinfo {title} {Nonresonant inelastic x-ray scattering
  involving excitonic excitations: The examples of {NiO} and {CoO}},\ }\href
  {https://doi.org/10.1103/PhysRevLett.99.257401} {\bibfield  {journal}
  {\bibinfo  {journal} {Phys. Rev. Lett.}\ }\textbf {\bibinfo {volume} {99}},\
  \bibinfo {pages} {257401} (\bibinfo {year} {2007})}\BibitemShut {NoStop}%
\bibitem [{\citenamefont {Gordon}\ \emph {et~al.}(2008)\citenamefont {Gordon},
  \citenamefont {Seidler}, \citenamefont {Fister}, \citenamefont {Haverkort},
  \citenamefont {Sawatzky}, \citenamefont {Tanaka},\ and\ \citenamefont
  {Sham}}]{Gordon2008}%
  \BibitemOpen
  \bibfield  {author} {\bibinfo {author} {\bibfnamefont {R.~A.}\ \bibnamefont
  {Gordon}}, \bibinfo {author} {\bibfnamefont {G.~T.}\ \bibnamefont {Seidler}},
  \bibinfo {author} {\bibfnamefont {T.~T.}\ \bibnamefont {Fister}}, \bibinfo
  {author} {\bibfnamefont {M.~W.}\ \bibnamefont {Haverkort}}, \bibinfo {author}
  {\bibfnamefont {G.~A.}\ \bibnamefont {Sawatzky}}, \bibinfo {author}
  {\bibfnamefont {A.}~\bibnamefont {Tanaka}},\ and\ \bibinfo {author}
  {\bibfnamefont {T.~K.}\ \bibnamefont {Sham}},\ }\bibfield  {title} {\bibinfo
  {title} {High multipole transitions in {NIXS}: Valence and hybridization in
  4f systems},\ }\href {https://doi.org/10.1209/0295-5075/81/26004} {\bibfield
  {journal} {\bibinfo  {journal} {EPL (Europhysics Letters)}\ }\textbf
  {\bibinfo {volume} {81}},\ \bibinfo {pages} {26004} (\bibinfo {year}
  {2008})}\BibitemShut {NoStop}%
\bibitem [{\citenamefont {Gordon}\ \emph {et~al.}(2009)\citenamefont {Gordon},
  \citenamefont {Haverkort}, \citenamefont {Sen~Gupta},\ and\ \citenamefont
  {G.A.}}]{Gordon2009}%
  \BibitemOpen
  \bibfield  {author} {\bibinfo {author} {\bibfnamefont {R.}~\bibnamefont
  {Gordon}}, \bibinfo {author} {\bibfnamefont {M.}~\bibnamefont {Haverkort}},
  \bibinfo {author} {\bibfnamefont {S.}~\bibnamefont {Sen~Gupta}},\ and\
  \bibinfo {author} {\bibfnamefont {S.}~\bibnamefont {G.A.}},\ }\bibfield
  {title} {\bibinfo {title} {Orientation-dependent x-ray {R}aman scattering
  from cubic crystals: {N}atural linear dichroism in{ MnO} and {CeO}$_2$.},\
  }\href {https://doi.org/10.1088/1742-6596/190/1/012047} {\bibfield  {journal}
  {\bibinfo  {journal} {J. Phys. Conf. Ser.}\ }\textbf {\bibinfo {volume}
  {190}},\ \bibinfo {pages} {012047} (\bibinfo {year} {2009})}\BibitemShut
  {NoStop}%
\bibitem [{\citenamefont {Willers}\ \emph
  {et~al.}(2012{\natexlab{a}})\citenamefont {Willers}, \citenamefont
  {Strigari}, \citenamefont {Hiraoka}, \citenamefont {Cai}, \citenamefont
  {Haverkort}, \citenamefont {Tsuei}, \citenamefont {Liao}, \citenamefont
  {Seiro}, \citenamefont {Geibel}, \citenamefont {Steglich}, \citenamefont
  {Tjeng},\ and\ \citenamefont {Severing}}]{Willers2012a}%
  \BibitemOpen
  \bibfield  {author} {\bibinfo {author} {\bibfnamefont {T.}~\bibnamefont
  {Willers}}, \bibinfo {author} {\bibfnamefont {F.}~\bibnamefont {Strigari}},
  \bibinfo {author} {\bibfnamefont {N.}~\bibnamefont {Hiraoka}}, \bibinfo
  {author} {\bibfnamefont {Y.~Q.}\ \bibnamefont {Cai}}, \bibinfo {author}
  {\bibfnamefont {M.~W.}\ \bibnamefont {Haverkort}}, \bibinfo {author}
  {\bibfnamefont {K.-D.}\ \bibnamefont {Tsuei}}, \bibinfo {author}
  {\bibfnamefont {Y.~F.}\ \bibnamefont {Liao}}, \bibinfo {author}
  {\bibfnamefont {S.}~\bibnamefont {Seiro}}, \bibinfo {author} {\bibfnamefont
  {C.}~\bibnamefont {Geibel}}, \bibinfo {author} {\bibfnamefont
  {F.}~\bibnamefont {Steglich}}, \bibinfo {author} {\bibfnamefont {L.~H.}\
  \bibnamefont {Tjeng}},\ and\ \bibinfo {author} {\bibfnamefont
  {A.}~\bibnamefont {Severing}},\ }\bibfield  {title} {\bibinfo {title}
  {Determining the in-plane orientation of the ground-state orbital of
  {CeCu}$_2${Si}$_2$},\ }\href@noop {} {\bibfield  {journal} {\bibinfo
  {journal} {Phys. Rev. Lett.}\ }\textbf {\bibinfo {volume} {109}},\ \bibinfo
  {pages} {046401} (\bibinfo {year} {2012}{\natexlab{a}})},\ \bibinfo {note}
  {note: Fig.\,1 of Ref.\,\cite{Willers2012a} used
  $\left|\alpha\right|$\,=\,0.88 as proposed by Goremychkin while
  Fig.\,\ref{unit_cell} in this work uses the $\left|\alpha\right|$ value as
  determined in the present XAS experiment.}\BibitemShut {Stop}%
\bibitem [{\citenamefont {Tanaka}\ and\ \citenamefont {Jo}(1994)}]{Tanaka1994}%
  \BibitemOpen
  \bibfield  {author} {\bibinfo {author} {\bibfnamefont {A.}~\bibnamefont
  {Tanaka}}\ and\ \bibinfo {author} {\bibfnamefont {T.}~\bibnamefont {Jo}},\
  }\bibfield  {title} {\bibinfo {title} {Resonant 3d, 3p and 3s photoemission
  in transition metal oxides predicted at 2p threshold},\ }\href
  {https://doi.org/10.1143/JPSJ.63.2788} {\bibfield  {journal} {\bibinfo
  {journal} {J Phys. Soc. Jpn.}\ }\textbf {\bibinfo {volume} {63}},\ \bibinfo
  {pages} {2788} (\bibinfo {year} {1994})}\BibitemShut {NoStop}%
\bibitem [{Tho(1997)}]{Thole1997}%
  \BibitemOpen
  \href@noop {} {\emph {\bibinfo {title} {Theo Thole Memorial Issue}}}\
  (\bibinfo  {publisher} {J. Electron Spectroscopy Relat. Phenom. 86 (1-3), pp
  1-207},\ \bibinfo {year} {1997})\BibitemShut {NoStop}%
\bibitem [{\citenamefont {de~Groot}\ and\ \citenamefont
  {Kotani}(2008)}]{deGroot2008}%
  \BibitemOpen
  \bibfield  {author} {\bibinfo {author} {\bibfnamefont {F.}~\bibnamefont
  {de~Groot}}\ and\ \bibinfo {author} {\bibfnamefont {A.}~\bibnamefont
  {Kotani}},\ }\href@noop {} {\emph {\bibinfo {title} {Core Level Spectroscopy
  of Solids}}}\ (\bibinfo  {publisher} {Taylor and Francis Group, LLC},\
  \bibinfo {year} {2008})\BibitemShut {NoStop}%
\bibitem [{\citenamefont {Hansmann}\ \emph {et~al.}(2008)\citenamefont
  {Hansmann}, \citenamefont {Severing}, \citenamefont {Hu}, \citenamefont
  {Haverkort}, \citenamefont {Chang}, \citenamefont {Klein}, \citenamefont
  {Tanaka}, \citenamefont {Hsieh}, \citenamefont {Lin}, \citenamefont {Chen},
  \citenamefont {F\aa~k}, \citenamefont {Lejay},\ and\ \citenamefont
  {Tjeng}}]{Hansmann2008}%
  \BibitemOpen
  \bibfield  {author} {\bibinfo {author} {\bibfnamefont {P.}~\bibnamefont
  {Hansmann}}, \bibinfo {author} {\bibfnamefont {A.}~\bibnamefont {Severing}},
  \bibinfo {author} {\bibfnamefont {Z.}~\bibnamefont {Hu}}, \bibinfo {author}
  {\bibfnamefont {M.~W.}\ \bibnamefont {Haverkort}}, \bibinfo {author}
  {\bibfnamefont {C.~F.}\ \bibnamefont {Chang}}, \bibinfo {author}
  {\bibfnamefont {S.}~\bibnamefont {Klein}}, \bibinfo {author} {\bibfnamefont
  {A.}~\bibnamefont {Tanaka}}, \bibinfo {author} {\bibfnamefont {H.~H.}\
  \bibnamefont {Hsieh}}, \bibinfo {author} {\bibfnamefont {H.-J.}\ \bibnamefont
  {Lin}}, \bibinfo {author} {\bibfnamefont {C.~T.}\ \bibnamefont {Chen}},
  \bibinfo {author} {\bibfnamefont {B.}~\bibnamefont {F\aa~k}}, \bibinfo
  {author} {\bibfnamefont {P.}~\bibnamefont {Lejay}},\ and\ \bibinfo {author}
  {\bibfnamefont {L.~H.}\ \bibnamefont {Tjeng}},\ }\bibfield  {title} {\bibinfo
  {title} {Determining the crystal-field ground state in rare earth heavy
  fermion materials using soft-x-ray absorption spectroscopy},\ }\href@noop {}
  {\bibfield  {journal} {\bibinfo  {journal} {Phys. Rev. Lett.}\ }\textbf
  {\bibinfo {volume} {100}},\ \bibinfo {pages} {066405} (\bibinfo {year}
  {2008})}\BibitemShut {NoStop}%
\bibitem [{\citenamefont {Willers}\ \emph {et~al.}(2010)\citenamefont
  {Willers}, \citenamefont {Hu}, \citenamefont {Hollmann}, \citenamefont
  {K\"orner}, \citenamefont {Gegner}, \citenamefont {Burnus}, \citenamefont
  {Fujiwara}, \citenamefont {Tanaka}, \citenamefont {Schmitz}, \citenamefont
  {Hsieh}, \citenamefont {Lin}, \citenamefont {Chen}, \citenamefont {Bauer},
  \citenamefont {Sarrao}, \citenamefont {Goremychkin}, \citenamefont {Koza},
  \citenamefont {Tjeng},\ and\ \citenamefont {Severing}}]{Willers2010}%
  \BibitemOpen
  \bibfield  {author} {\bibinfo {author} {\bibfnamefont {T.}~\bibnamefont
  {Willers}}, \bibinfo {author} {\bibfnamefont {Z.}~\bibnamefont {Hu}},
  \bibinfo {author} {\bibfnamefont {N.}~\bibnamefont {Hollmann}}, \bibinfo
  {author} {\bibfnamefont {P.~O.}\ \bibnamefont {K\"orner}}, \bibinfo {author}
  {\bibfnamefont {J.}~\bibnamefont {Gegner}}, \bibinfo {author} {\bibfnamefont
  {T.}~\bibnamefont {Burnus}}, \bibinfo {author} {\bibfnamefont
  {H.}~\bibnamefont {Fujiwara}}, \bibinfo {author} {\bibfnamefont
  {A.}~\bibnamefont {Tanaka}}, \bibinfo {author} {\bibfnamefont
  {D.}~\bibnamefont {Schmitz}}, \bibinfo {author} {\bibfnamefont {H.~H.}\
  \bibnamefont {Hsieh}}, \bibinfo {author} {\bibfnamefont {H.-J.}\ \bibnamefont
  {Lin}}, \bibinfo {author} {\bibfnamefont {C.~T.}\ \bibnamefont {Chen}},
  \bibinfo {author} {\bibfnamefont {E.~D.}\ \bibnamefont {Bauer}}, \bibinfo
  {author} {\bibfnamefont {J.~L.}\ \bibnamefont {Sarrao}}, \bibinfo {author}
  {\bibfnamefont {E.}~\bibnamefont {Goremychkin}}, \bibinfo {author}
  {\bibfnamefont {M.}~\bibnamefont {Koza}}, \bibinfo {author} {\bibfnamefont
  {L.~H.}\ \bibnamefont {Tjeng}},\ and\ \bibinfo {author} {\bibfnamefont
  {A.}~\bibnamefont {Severing}},\ }\bibfield  {title} {\bibinfo {title}
  {Crystal-field and kondo-scale investigations of $\text{Ce}m{\text{in}}_{5}$
  ($m=\text{Co}$, ir, and rh): A combined x-ray absorption and inelastic
  neutron scattering study},\ }\href
  {https://doi.org/10.1103/PhysRevB.81.195114} {\bibfield  {journal} {\bibinfo
  {journal} {Phys. Rev. B}\ }\textbf {\bibinfo {volume} {81}},\ \bibinfo
  {pages} {195114} (\bibinfo {year} {2010})}\BibitemShut {NoStop}%
\bibitem [{\citenamefont {Willers}\ \emph
  {et~al.}(2012{\natexlab{b}})\citenamefont {Willers}, \citenamefont {Adroja},
  \citenamefont {Rainford}, \citenamefont {Hu}, \citenamefont {Hollmann},
  \citenamefont {K\"orner}, \citenamefont {Chin}, \citenamefont {Schmitz},
  \citenamefont {Hsieh}, \citenamefont {Lin}, \citenamefont {Chen},
  \citenamefont {Bauer}, \citenamefont {Sarrao}, \citenamefont {McClellan},
  \citenamefont {Byler}, \citenamefont {Geibel}, \citenamefont {Steglich},
  \citenamefont {Aoki}, \citenamefont {Lejay}, \citenamefont {Tanaka},
  \citenamefont {Tjeng},\ and\ \citenamefont {Severing}}]{Willers2012}%
  \BibitemOpen
  \bibfield  {author} {\bibinfo {author} {\bibfnamefont {T.}~\bibnamefont
  {Willers}}, \bibinfo {author} {\bibfnamefont {D.~T.}\ \bibnamefont {Adroja}},
  \bibinfo {author} {\bibfnamefont {B.~D.}\ \bibnamefont {Rainford}}, \bibinfo
  {author} {\bibfnamefont {Z.}~\bibnamefont {Hu}}, \bibinfo {author}
  {\bibfnamefont {N.}~\bibnamefont {Hollmann}}, \bibinfo {author}
  {\bibfnamefont {P.~O.}\ \bibnamefont {K\"orner}}, \bibinfo {author}
  {\bibfnamefont {Y.-Y.}\ \bibnamefont {Chin}}, \bibinfo {author}
  {\bibfnamefont {D.}~\bibnamefont {Schmitz}}, \bibinfo {author} {\bibfnamefont
  {H.~H.}\ \bibnamefont {Hsieh}}, \bibinfo {author} {\bibfnamefont {H.-J.}\
  \bibnamefont {Lin}}, \bibinfo {author} {\bibfnamefont {C.~T.}\ \bibnamefont
  {Chen}}, \bibinfo {author} {\bibfnamefont {E.~D.}\ \bibnamefont {Bauer}},
  \bibinfo {author} {\bibfnamefont {J.~L.}\ \bibnamefont {Sarrao}}, \bibinfo
  {author} {\bibfnamefont {K.~J.}\ \bibnamefont {McClellan}}, \bibinfo {author}
  {\bibfnamefont {D.}~\bibnamefont {Byler}}, \bibinfo {author} {\bibfnamefont
  {C.}~\bibnamefont {Geibel}}, \bibinfo {author} {\bibfnamefont
  {F.}~\bibnamefont {Steglich}}, \bibinfo {author} {\bibfnamefont
  {H.}~\bibnamefont {Aoki}}, \bibinfo {author} {\bibfnamefont {P.}~\bibnamefont
  {Lejay}}, \bibinfo {author} {\bibfnamefont {A.}~\bibnamefont {Tanaka}},
  \bibinfo {author} {\bibfnamefont {L.~H.}\ \bibnamefont {Tjeng}},\ and\
  \bibinfo {author} {\bibfnamefont {A.}~\bibnamefont {Severing}},\ }\bibfield
  {title} {\bibinfo {title} {Spectroscopic determination of crystal-field
  levels in {CeRh$_2$Si$_2$} and {CeRu$_2$Si$_2$} and of the $4{f}^{0}$
  contributions in {Ce$M_2$Si$_2$} {($M$=Cu, Ru, Rh, Pd, and Au)}},\ }\href
  {https://doi.org/10.1103/PhysRevB.85.035117} {\bibfield  {journal} {\bibinfo
  {journal} {Phys. Rev. B}\ }\textbf {\bibinfo {volume} {85}},\ \bibinfo
  {pages} {035117} (\bibinfo {year} {2012}{\natexlab{b}})}\BibitemShut
  {NoStop}%
\bibitem [{\citenamefont {Strigari}\ \emph {et~al.}(2012)\citenamefont
  {Strigari}, \citenamefont {Willers}, \citenamefont {Muro}, \citenamefont
  {Yutani}, \citenamefont {Takabatake}, \citenamefont {Hu}, \citenamefont
  {Chin}, \citenamefont {Agrestini}, \citenamefont {Lin}, \citenamefont {Chen},
  \citenamefont {Tanaka}, \citenamefont {Haverkort}, \citenamefont {Tjeng},\
  and\ \citenamefont {Severing}}]{Strigari2012}%
  \BibitemOpen
  \bibfield  {author} {\bibinfo {author} {\bibfnamefont {F.}~\bibnamefont
  {Strigari}}, \bibinfo {author} {\bibfnamefont {T.}~\bibnamefont {Willers}},
  \bibinfo {author} {\bibfnamefont {Y.}~\bibnamefont {Muro}}, \bibinfo {author}
  {\bibfnamefont {K.}~\bibnamefont {Yutani}}, \bibinfo {author} {\bibfnamefont
  {T.}~\bibnamefont {Takabatake}}, \bibinfo {author} {\bibfnamefont
  {Z.}~\bibnamefont {Hu}}, \bibinfo {author} {\bibfnamefont {Y.-Y.}\
  \bibnamefont {Chin}}, \bibinfo {author} {\bibfnamefont {S.}~\bibnamefont
  {Agrestini}}, \bibinfo {author} {\bibfnamefont {H.-J.}\ \bibnamefont {Lin}},
  \bibinfo {author} {\bibfnamefont {C.~T.}\ \bibnamefont {Chen}}, \bibinfo
  {author} {\bibfnamefont {A.}~\bibnamefont {Tanaka}}, \bibinfo {author}
  {\bibfnamefont {M.~W.}\ \bibnamefont {Haverkort}}, \bibinfo {author}
  {\bibfnamefont {L.~H.}\ \bibnamefont {Tjeng}},\ and\ \bibinfo {author}
  {\bibfnamefont {A.}~\bibnamefont {Severing}},\ }\bibfield  {title} {\bibinfo
  {title} {Crystal-field ground state of the orthorhombic {K}ondo insulator
  {CeRu}$_2${Al}$_{10}$},\ }\href@noop {} {\bibfield  {journal} {\bibinfo
  {journal} {Phys. Rev. B}\ }\textbf {\bibinfo {volume} {86}},\ \bibinfo
  {pages} {081105} (\bibinfo {year} {2012})}\BibitemShut {NoStop}%
\bibitem [{\citenamefont {Willers}\ \emph {et~al.}(2015)\citenamefont
  {Willers}, \citenamefont {Strigari}, \citenamefont {Hu}, \citenamefont
  {Sessi}, \citenamefont {Brookes}, \citenamefont {Bauer}, \citenamefont
  {Sarrao}, \citenamefont {Thompson}, \citenamefont {Tanaka}, \citenamefont
  {Wirth}, \citenamefont {Tjeng},\ and\ \citenamefont
  {Severing}}]{Willers2015}%
  \BibitemOpen
  \bibfield  {author} {\bibinfo {author} {\bibfnamefont {T.}~\bibnamefont
  {Willers}}, \bibinfo {author} {\bibfnamefont {F.}~\bibnamefont {Strigari}},
  \bibinfo {author} {\bibfnamefont {Z.}~\bibnamefont {Hu}}, \bibinfo {author}
  {\bibfnamefont {V.}~\bibnamefont {Sessi}}, \bibinfo {author} {\bibfnamefont
  {N.}~\bibnamefont {Brookes}}, \bibinfo {author} {\bibfnamefont
  {E.}~\bibnamefont {Bauer}}, \bibinfo {author} {\bibfnamefont
  {J.}~\bibnamefont {Sarrao}}, \bibinfo {author} {\bibfnamefont
  {J.}~\bibnamefont {Thompson}}, \bibinfo {author} {\bibfnamefont
  {A.}~\bibnamefont {Tanaka}}, \bibinfo {author} {\bibfnamefont
  {S.}~\bibnamefont {Wirth}}, \bibinfo {author} {\bibfnamefont
  {L.}~\bibnamefont {Tjeng}},\ and\ \bibinfo {author} {\bibfnamefont
  {A.}~\bibnamefont {Severing}},\ }\bibfield  {title} {\bibinfo {title}
  {Correlation between ground state and orbital anisotropy in heavy fermion
  materials},\ }\href {https://doi.org/10.1073/pnas.1415657112} {\bibfield
  {journal} {\bibinfo  {journal} {Proc. Nat. Acad. Science U.S.A.}\ }\textbf
  {\bibinfo {volume} {112}},\ \bibinfo {pages} {2384} (\bibinfo {year}
  {2015})}\BibitemShut {NoStop}%
\bibitem [{\citenamefont {Gunnarsson}\ and\ \citenamefont
  {Sch\"onhammer}(1983)}]{Gunnarsson1983}%
  \BibitemOpen
  \bibfield  {author} {\bibinfo {author} {\bibfnamefont {O.}~\bibnamefont
  {Gunnarsson}}\ and\ \bibinfo {author} {\bibfnamefont {K.}~\bibnamefont
  {Sch\"onhammer}},\ }\bibfield  {title} {\bibinfo {title} {Electron
  spectroscopies for ce compounds in the impurity model},\ }\href
  {https://doi.org/10.1103/PhysRevB.28.4315} {\bibfield  {journal} {\bibinfo
  {journal} {Phys. Rev. B}\ }\textbf {\bibinfo {volume} {28}},\ \bibinfo
  {pages} {4315} (\bibinfo {year} {1983})}\BibitemShut {NoStop}%
\bibitem [{\citenamefont {Seiro}\ \emph {et~al.}(2010)\citenamefont {Seiro},
  \citenamefont {Deppe}, \citenamefont {Jeevan}, \citenamefont {Burkhardt},\
  and\ \citenamefont {Geibel}}]{Seiro2010}%
  \BibitemOpen
  \bibfield  {author} {\bibinfo {author} {\bibfnamefont {S.}~\bibnamefont
  {Seiro}}, \bibinfo {author} {\bibfnamefont {M.}~\bibnamefont {Deppe}},
  \bibinfo {author} {\bibfnamefont {H.}~\bibnamefont {Jeevan}}, \bibinfo
  {author} {\bibfnamefont {U.}~\bibnamefont {Burkhardt}},\ and\ \bibinfo
  {author} {\bibfnamefont {C.}~\bibnamefont {Geibel}},\ }\bibfield  {title}
  {\bibinfo {title} {Flux crystal growth of {CeCu$_2$Si$_2$}: Revealing the
  effect of composition},\ }\href {https://doi.org/10.1002/pssb.200983039}
  {\bibfield  {journal} {\bibinfo  {journal} {physica status solidi (b)}\
  }\textbf {\bibinfo {volume} {247}},\ \bibinfo {pages} {614} (\bibinfo {year}
  {2010})}\BibitemShut {NoStop}%
\bibitem [{\citenamefont {Joly}\ \emph {et~al.}(2014)\citenamefont {Joly},
  \citenamefont {Otero}, \citenamefont {Choueikani}, \citenamefont {Marteau},
  \citenamefont {Chapuis},\ and\ \citenamefont {Ohresser}}]{Joly2014}%
  \BibitemOpen
  \bibfield  {author} {\bibinfo {author} {\bibfnamefont {L.}~\bibnamefont
  {Joly}}, \bibinfo {author} {\bibfnamefont {E.}~\bibnamefont {Otero}},
  \bibinfo {author} {\bibfnamefont {F.}~\bibnamefont {Choueikani}}, \bibinfo
  {author} {\bibfnamefont {F.}~\bibnamefont {Marteau}}, \bibinfo {author}
  {\bibfnamefont {L.}~\bibnamefont {Chapuis}},\ and\ \bibinfo {author}
  {\bibfnamefont {P.}~\bibnamefont {Ohresser}},\ }\bibfield  {title} {\bibinfo
  {title} {{Fast continuous energy scan with dynamic coupling of the
  monochromator and undulator at the DEIMOS beamline}},\ }\href
  {https://doi.org/10.1107/S1600577514003671} {\bibfield  {journal} {\bibinfo
  {journal} {J. Synchrotron Rad.}\ }\textbf {\bibinfo {volume} {21}},\ \bibinfo
  {pages} {502} (\bibinfo {year} {2014})}\BibitemShut {NoStop}%
\bibitem [{\citenamefont {Barla}\ \emph {et~al.}(2016)\citenamefont {Barla},
  \citenamefont {Nicolas}, \citenamefont {Cocco}, \citenamefont {Valvidares},
  \citenamefont {Herrero-Martin}, \citenamefont {Gargiani}, \citenamefont
  {Moldes}, \citenamefont {Ruget}, \citenamefont {Pellegrin},\ and\
  \citenamefont {S.}}]{Barla2016}%
  \BibitemOpen
  \bibfield  {author} {\bibinfo {author} {\bibfnamefont {A.}~\bibnamefont
  {Barla}}, \bibinfo {author} {\bibfnamefont {J.}~\bibnamefont {Nicolas}},
  \bibinfo {author} {\bibfnamefont {D.}~\bibnamefont {Cocco}}, \bibinfo
  {author} {\bibfnamefont {S.}~\bibnamefont {Valvidares}}, \bibinfo {author}
  {\bibfnamefont {J.}~\bibnamefont {Herrero-Martin}}, \bibinfo {author}
  {\bibfnamefont {P.}~\bibnamefont {Gargiani}}, \bibinfo {author}
  {\bibfnamefont {J.}~\bibnamefont {Moldes}}, \bibinfo {author} {\bibfnamefont
  {C.}~\bibnamefont {Ruget}}, \bibinfo {author} {\bibfnamefont
  {E.}~\bibnamefont {Pellegrin}},\ and\ \bibinfo {author} {\bibfnamefont
  {F.}~\bibnamefont {S.}},\ }\bibfield  {title} {\bibinfo {title} {Design and
  performance of {BOREAS}, the beamline for resonant x-ray absorption and
  scattering experiments at the {ALBA} synchrotron light source},\ }\href
  {https://doi.org/10.1107/S1600577516013461} {\bibfield  {journal} {\bibinfo
  {journal} {J. Synchrotron Rad.}\ }\textbf {\bibinfo {volume} {23}},\ \bibinfo
  {pages} {1507} (\bibinfo {year} {2016})}\BibitemShut {NoStop}%
\bibitem [{\citenamefont {Kappler}\ \emph {et~al.}(2018)\citenamefont
  {Kappler}, \citenamefont {Otero}, \citenamefont {Li}, \citenamefont {Joly},
  \citenamefont {Schmerber}, \citenamefont {Muller}, \citenamefont {Scheurer},
  \citenamefont {Leduc}, \citenamefont {Gobaut}, \citenamefont {Poggini},
  \citenamefont {Serrano}, \citenamefont {Choueikani}, \citenamefont {Lhotel},
  \citenamefont {Cornia}, \citenamefont {Sessoli}, \citenamefont {Mannini},
  \citenamefont {Arrio}, \citenamefont {Sainctavit},\ and\ \citenamefont
  {Ohresser}}]{Kappler2018}%
  \BibitemOpen
  \bibfield  {author} {\bibinfo {author} {\bibfnamefont {J.-P.}\ \bibnamefont
  {Kappler}}, \bibinfo {author} {\bibfnamefont {E.}~\bibnamefont {Otero}},
  \bibinfo {author} {\bibfnamefont {W.}~\bibnamefont {Li}}, \bibinfo {author}
  {\bibfnamefont {L.}~\bibnamefont {Joly}}, \bibinfo {author} {\bibfnamefont
  {G.}~\bibnamefont {Schmerber}}, \bibinfo {author} {\bibfnamefont
  {B.}~\bibnamefont {Muller}}, \bibinfo {author} {\bibfnamefont
  {F.}~\bibnamefont {Scheurer}}, \bibinfo {author} {\bibfnamefont
  {F.}~\bibnamefont {Leduc}}, \bibinfo {author} {\bibfnamefont
  {B.}~\bibnamefont {Gobaut}}, \bibinfo {author} {\bibfnamefont
  {L.}~\bibnamefont {Poggini}}, \bibinfo {author} {\bibfnamefont
  {G.}~\bibnamefont {Serrano}}, \bibinfo {author} {\bibfnamefont
  {F.}~\bibnamefont {Choueikani}}, \bibinfo {author} {\bibfnamefont
  {E.}~\bibnamefont {Lhotel}}, \bibinfo {author} {\bibfnamefont
  {A.}~\bibnamefont {Cornia}}, \bibinfo {author} {\bibfnamefont
  {R.}~\bibnamefont {Sessoli}}, \bibinfo {author} {\bibfnamefont
  {M.}~\bibnamefont {Mannini}}, \bibinfo {author} {\bibfnamefont {M.-A.}\
  \bibnamefont {Arrio}}, \bibinfo {author} {\bibfnamefont {P.}~\bibnamefont
  {Sainctavit}},\ and\ \bibinfo {author} {\bibfnamefont {P.}~\bibnamefont
  {Ohresser}},\ }\bibfield  {title} {\bibinfo {title} {{Ultralow-temperature
  device dedicated to soft X-ray magnetic circular dichroism experiments}},\
  }\href {https://doi.org/10.1107/S1600577518012717} {\bibfield  {journal}
  {\bibinfo  {journal} {J. Synchrotron Rad.}\ }\textbf {\bibinfo {volume}
  {25}},\ \bibinfo {pages} {1727} (\bibinfo {year} {2018})}\BibitemShut
  {NoStop}%
\bibitem [{\citenamefont {Haverkort}(2016)}]{Haverkort2016}%
  \BibitemOpen
  \bibfield  {author} {\bibinfo {author} {\bibfnamefont {M.~W.}\ \bibnamefont
  {Haverkort}},\ }\bibfield  {title} {\bibinfo {title} {${Q}uanty$ for core
  level spectroscopy - excitons, resonances and band excitations in time and
  frequency domain},\ }\href {https://doi.org/10.1088/1742-6596/712/1/012001}
  {\bibfield  {journal} {\bibinfo  {journal} {J. Phys.: Conf. Ser.}\ }\textbf
  {\bibinfo {volume} {712}},\ \bibinfo {pages} {012001} (\bibinfo {year}
  {2016})}\BibitemShut {NoStop}%
\bibitem [{\citenamefont {Cowan}(1981)}]{Cowan1981}%
  \BibitemOpen
  \bibfield  {author} {\bibinfo {author} {\bibfnamefont {R.}~\bibnamefont
  {Cowan}},\ }\href@noop {} {\emph {\bibinfo {title} {The theory of atomic
  structure and spectra.}}}\ (\bibinfo  {publisher} {University of California
  Press},\ \bibinfo {year} {1981})\BibitemShut {NoStop}%
\bibitem [{\citenamefont {Horn}\ \emph {et~al.}(1981)\citenamefont {Horn},
  \citenamefont {Holland-Moritz}, \citenamefont {Loewenhaupt}, \citenamefont
  {Steglich}, \citenamefont {Scheuer}, \citenamefont {Benoit},\ and\
  \citenamefont {Flouquet}}]{Horn1981}%
  \BibitemOpen
  \bibfield  {author} {\bibinfo {author} {\bibfnamefont {S.}~\bibnamefont
  {Horn}}, \bibinfo {author} {\bibfnamefont {E.}~\bibnamefont
  {Holland-Moritz}}, \bibinfo {author} {\bibfnamefont {M.}~\bibnamefont
  {Loewenhaupt}}, \bibinfo {author} {\bibfnamefont {F.}~\bibnamefont
  {Steglich}}, \bibinfo {author} {\bibfnamefont {H.}~\bibnamefont {Scheuer}},
  \bibinfo {author} {\bibfnamefont {A.}~\bibnamefont {Benoit}},\ and\ \bibinfo
  {author} {\bibfnamefont {J.}~\bibnamefont {Flouquet}},\ }\bibfield  {title}
  {\bibinfo {title} {Magnetic neutron scattering and crystal-field states in
  {CeCu}$_2${Si}$_2$},\ }\href {https://doi.org/10.1103/PhysRevB.23.3171}
  {\bibfield  {journal} {\bibinfo  {journal} {Phys. Rev. B}\ }\textbf {\bibinfo
  {volume} {23}},\ \bibinfo {pages} {3171} (\bibinfo {year}
  {1981})}\BibitemShut {NoStop}%
\bibitem [{\citenamefont {Rueff}\ \emph {et~al.}(2015)\citenamefont {Rueff},
  \citenamefont {Ablett}, \citenamefont {Strigari}, \citenamefont {Deppe},
  \citenamefont {Haverkort}, \citenamefont {Tjeng},\ and\ \citenamefont
  {Severing}}]{Rueff2015}%
  \BibitemOpen
  \bibfield  {author} {\bibinfo {author} {\bibfnamefont {J.-P.}\ \bibnamefont
  {Rueff}}, \bibinfo {author} {\bibfnamefont {J.~M.}\ \bibnamefont {Ablett}},
  \bibinfo {author} {\bibfnamefont {F.}~\bibnamefont {Strigari}}, \bibinfo
  {author} {\bibfnamefont {M.}~\bibnamefont {Deppe}}, \bibinfo {author}
  {\bibfnamefont {M.~W.}\ \bibnamefont {Haverkort}}, \bibinfo {author}
  {\bibfnamefont {L.~H.}\ \bibnamefont {Tjeng}},\ and\ \bibinfo {author}
  {\bibfnamefont {A.}~\bibnamefont {Severing}},\ }\bibfield  {title} {\bibinfo
  {title} {Absence of orbital rotation in superconducting {CeCu}$_2${Ge}$_2$},\
  }\href {https://doi.org/10.1103/PhysRevB.91.201108} {\bibfield  {journal}
  {\bibinfo  {journal} {Phys. Rev. B}\ }\textbf {\bibinfo {volume} {91}},\
  \bibinfo {pages} {201108} (\bibinfo {year} {2015})}\BibitemShut {NoStop}%
\bibitem [{\citenamefont {Sundermann}\ \emph {et~al.}(2015)\citenamefont
  {Sundermann}, \citenamefont {Strigari}, \citenamefont {Willers},
  \citenamefont {Winkler}, \citenamefont {Prokofiev}, \citenamefont
  {A.~Ablett}, \citenamefont {Rueff}, \citenamefont {Schmitz}, \citenamefont
  {Weschke}, \citenamefont {Sala-Moretti}, \citenamefont {Al-Zein},
  \citenamefont {Tanaka}, \citenamefont {Haverkort}, \citenamefont
  {Kasinathan}, \citenamefont {Tjeng}, \citenamefont {Paschen},\ and\
  \citenamefont {Severing}}]{Sundermann2015}%
  \BibitemOpen
  \bibfield  {author} {\bibinfo {author} {\bibfnamefont {M.}~\bibnamefont
  {Sundermann}}, \bibinfo {author} {\bibfnamefont {F.}~\bibnamefont
  {Strigari}}, \bibinfo {author} {\bibfnamefont {T.}~\bibnamefont {Willers}},
  \bibinfo {author} {\bibfnamefont {H.}~\bibnamefont {Winkler}}, \bibinfo
  {author} {\bibnamefont {Prokofiev}}, \bibinfo {author} {\bibfnamefont
  {J.~M.}\ \bibnamefont {A.~Ablett}}, \bibinfo {author} {\bibfnamefont {J.-P.}\
  \bibnamefont {Rueff}}, \bibinfo {author} {\bibfnamefont {D.}~\bibnamefont
  {Schmitz}}, \bibinfo {author} {\bibfnamefont {E.}~\bibnamefont {Weschke}},
  \bibinfo {author} {\bibfnamefont {M.}~\bibnamefont {Sala-Moretti}}, \bibinfo
  {author} {\bibfnamefont {A.}~\bibnamefont {Al-Zein}}, \bibinfo {author}
  {\bibfnamefont {A.}~\bibnamefont {Tanaka}}, \bibinfo {author} {\bibfnamefont
  {M.~W.}\ \bibnamefont {Haverkort}}, \bibinfo {author} {\bibfnamefont
  {D.}~\bibnamefont {Kasinathan}}, \bibinfo {author} {\bibfnamefont {L.~H.}\
  \bibnamefont {Tjeng}}, \bibinfo {author} {\bibfnamefont {S.}~\bibnamefont
  {Paschen}},\ and\ \bibinfo {author} {\bibfnamefont {A.}~\bibnamefont
  {Severing}},\ }\bibfield  {title} {\bibinfo {title} {{CeRu$_4$Sn$_6$}: a
  strongly correlated material with nontrivial topology},\ }\href
  {https://doi.org/10.1038/srep17937} {\bibfield  {journal} {\bibinfo
  {journal} {Scientific Reports}\ }\textbf {\bibinfo {volume} {5}},\ \bibinfo
  {pages} {17937} (\bibinfo {year} {2015})}\BibitemShut {NoStop}%
\bibitem [{\citenamefont {Sundermann}\ \emph {et~al.}(2019)\citenamefont
  {Sundermann}, \citenamefont {Amorese}, \citenamefont {Strigari},
  \citenamefont {Leedahl}, \citenamefont {Tjeng}, \citenamefont {Haverkort},
  \citenamefont {Gretarsson}, \citenamefont {Yavas}, \citenamefont
  {Sala-Moretti}, \citenamefont {Bauer}, \citenamefont {Rosa}, \citenamefont
  {Thompson},\ and\ \citenamefont {Severing}}]{Sundermann2019}%
  \BibitemOpen
  \bibfield  {author} {\bibinfo {author} {\bibfnamefont {M.}~\bibnamefont
  {Sundermann}}, \bibinfo {author} {\bibfnamefont {A.}~\bibnamefont {Amorese}},
  \bibinfo {author} {\bibfnamefont {F.}~\bibnamefont {Strigari}}, \bibinfo
  {author} {\bibfnamefont {B.}~\bibnamefont {Leedahl}}, \bibinfo {author}
  {\bibfnamefont {L.~H.}\ \bibnamefont {Tjeng}}, \bibinfo {author}
  {\bibfnamefont {M.~W.}\ \bibnamefont {Haverkort}}, \bibinfo {author}
  {\bibfnamefont {H.}~\bibnamefont {Gretarsson}}, \bibinfo {author}
  {\bibfnamefont {H.}~\bibnamefont {Yavas}}, \bibinfo {author} {\bibfnamefont
  {M.}~\bibnamefont {Sala-Moretti}}, \bibinfo {author} {\bibfnamefont {E.~D.}\
  \bibnamefont {Bauer}}, \bibinfo {author} {\bibfnamefont {P.~F.~S.}\
  \bibnamefont {Rosa}}, \bibinfo {author} {\bibfnamefont {J.~D.}\ \bibnamefont
  {Thompson}},\ and\ \bibinfo {author} {\bibfnamefont {A.}~\bibnamefont
  {Severing}},\ }\bibfield  {title} {\bibinfo {title} {Orientation of the
  ground-state orbital in {CeCoIn}$_5$ and {CeRhIn}$_5$},\ }\href
  {https://doi.org/10.1103/PhysRevB.99.235143} {\bibfield  {journal} {\bibinfo
  {journal} {Phys. Rev. B}\ }\textbf {\bibinfo {volume} {99}},\ \bibinfo
  {pages} {235143} (\bibinfo {year} {2019})}\BibitemShut {NoStop}%
\bibitem [{\citenamefont {Bickers}\ \emph {et~al.}(1987)\citenamefont
  {Bickers}, \citenamefont {Cox},\ and\ \citenamefont {Wilkins}}]{Cox1987}%
  \BibitemOpen
  \bibfield  {author} {\bibinfo {author} {\bibfnamefont {N.~E.}\ \bibnamefont
  {Bickers}}, \bibinfo {author} {\bibfnamefont {D.~L.}\ \bibnamefont {Cox}},\
  and\ \bibinfo {author} {\bibfnamefont {J.~W.}\ \bibnamefont {Wilkins}},\
  }\bibfield  {title} {\bibinfo {title} {Self-consistent large-{N} expansion
  for normal-state properties of dilute magnetic alloys},\ }\href
  {https://doi.org/10.1103/PhysRevB.36.2036} {\bibfield  {journal} {\bibinfo
  {journal} {Phys. Rev. B}\ }\textbf {\bibinfo {volume} {36}},\ \bibinfo
  {pages} {2036} (\bibinfo {year} {1987})}\BibitemShut {NoStop}%
\bibitem [{\citenamefont {Kummer}\ \emph {et~al.}(2018)\citenamefont {Kummer},
  \citenamefont {Geibel}, \citenamefont {Krellner}, \citenamefont {Zwicknagl},
  \citenamefont {Laubschat}, \citenamefont {Brookes},\ and\ \citenamefont
  {Vyalikh}}]{Kummer2018}%
  \BibitemOpen
  \bibfield  {author} {\bibinfo {author} {\bibfnamefont {K.}~\bibnamefont
  {Kummer}}, \bibinfo {author} {\bibfnamefont {C.}~\bibnamefont {Geibel}},
  \bibinfo {author} {\bibfnamefont {C.}~\bibnamefont {Krellner}}, \bibinfo
  {author} {\bibfnamefont {G.}~\bibnamefont {Zwicknagl}}, \bibinfo {author}
  {\bibfnamefont {C.}~\bibnamefont {Laubschat}}, \bibinfo {author}
  {\bibfnamefont {N.~B.}\ \bibnamefont {Brookes}},\ and\ \bibinfo {author}
  {\bibfnamefont {D.~V.}\ \bibnamefont {Vyalikh}},\ }\bibfield  {title}
  {\bibinfo {title} {Similar temperature scale for valence changes in {K}ondo
  lattices with different {K}ondo temperatures},\ }\href
  {https://doi.org/10.1038/s41467-018-04438-8} {\bibfield  {journal} {\bibinfo
  {journal} {Nature Commun.}\ }\textbf {\bibinfo {volume} {9}},\ \bibinfo
  {pages} {2011} (\bibinfo {year} {2018})}\BibitemShut {NoStop}%
\bibitem [{\citenamefont {Zwicknagl}\ \emph {et~al.}(2010)\citenamefont
  {Zwicknagl}, \citenamefont {Zevin},\ and\ \citenamefont
  {Fulde}}]{Zwicknagl90a}%
  \BibitemOpen
  \bibfield  {author} {\bibinfo {author} {\bibfnamefont {G.}~\bibnamefont
  {Zwicknagl}}, \bibinfo {author} {\bibfnamefont {V.}~\bibnamefont {Zevin}},\
  and\ \bibinfo {author} {\bibfnamefont {P.}~\bibnamefont {Fulde}},\ }\bibfield
   {title} {\bibinfo {title} {Simple approximation scheme for the {A}nderson
  impurity {H}amiltonian},\ }\href {https://doi.org/10.1007/BF01437646}
  {\bibfield  {journal} {\bibinfo  {journal} {Z. Physik B}\ }\textbf {\bibinfo
  {volume} {79}},\ \bibinfo {pages} {365} (\bibinfo {year} {2010})}\BibitemShut
  {NoStop}%
\bibitem [{\citenamefont {Varma}\ and\ \citenamefont
  {Yafet}(1976)}]{Varma1976}%
  \BibitemOpen
  \bibfield  {author} {\bibinfo {author} {\bibfnamefont {C.~M.}\ \bibnamefont
  {Varma}}\ and\ \bibinfo {author} {\bibfnamefont {Y.}~\bibnamefont {Yafet}},\
  }\bibfield  {title} {\bibinfo {title} {Magnetic susceptibility of
  mixed-valence rare-earth compounds},\ }\href
  {https://doi.org/10.1103/PhysRevB.13.2950} {\bibfield  {journal} {\bibinfo
  {journal} {Phys. Rev. B}\ }\textbf {\bibinfo {volume} {13}},\ \bibinfo
  {pages} {2950} (\bibinfo {year} {1976})}\BibitemShut {NoStop}%
\bibitem [{\citenamefont {Zwicknagl}(1992)}]{Zwicknagl1992}%
  \BibitemOpen
  \bibfield  {author} {\bibinfo {author} {\bibfnamefont {G.}~\bibnamefont
  {Zwicknagl}},\ }\bibfield  {title} {\bibinfo {title} {Quasi-particles in
  heavy fermion systems},\ }\href {https://doi.org/10.1080/00018739200101503}
  {\bibfield  {journal} {\bibinfo  {journal} {Advances in Physics}\ }\textbf
  {\bibinfo {volume} {41}},\ \bibinfo {pages} {203} (\bibinfo {year} {1992})},\
  \Eprint {https://arxiv.org/abs/https://doi.org/10.1080/00018739200101503}
  {https://doi.org/10.1080/00018739200101503} \BibitemShut {NoStop}%
\bibitem [{\citenamefont {Zwicknagl}\ and\ \citenamefont
  {Pulst}(1993)}]{Zwick1993}%
  \BibitemOpen
  \bibfield  {author} {\bibinfo {author} {\bibfnamefont {G.}~\bibnamefont
  {Zwicknagl}}\ and\ \bibinfo {author} {\bibfnamefont {U.}~\bibnamefont
  {Pulst}},\ }\bibfield  {title} {\bibinfo {title} {Cecu$_2$si$_2$:
  Renormalized band structure, quasiparticles and co-operative phenomena},\
  }\href {https://doi.org/https://doi.org/10.1016/0921-4526(93)90736-P}
  {\bibfield  {journal} {\bibinfo  {journal} {Physica B: Condensed Matter}\
  }\textbf {\bibinfo {volume} {186-188}},\ \bibinfo {pages} {895 } (\bibinfo
  {year} {1993})}\BibitemShut {NoStop}%
\bibitem [{\citenamefont {Si}(2020)}]{Priv2020}%
  \BibitemOpen
  \bibfield  {author} {\bibinfo {author} {\bibfnamefont {Q.}~\bibnamefont
  {Si}},\ }\bibfield  {title} {\bibinfo {title} {private communication},\
  }\href@noop {} {\  (\bibinfo {year} {2020})}\BibitemShut {NoStop}%
\end{thebibliography}
\end{document}